\documentclass[letterpaper,twocolumn,table,xcdraw,dvipsnames,10pt]{article}
\usepackage{hyperref}
\usepackage{usenix2019_v3}

\pdfoutput=1

\usepackage{xspace}
\usepackage{xcolor}
\usepackage{longtable}
\usepackage{multirow}
\usepackage{multicol} %
\usepackage{comment}
\usepackage{enumerate}
\usepackage{graphicx}
\usepackage{tikz}
\usepackage{wasysym}
\usepackage{booktabs}
\usepackage{balance} %
\usepackage[inline]{enumitem} %
\usepackage{tabularx}
\usepackage{csquotes}
\usepackage{caption}
\usepackage{subcaption}
\usepackage{cleveref}
\usepackage{lscape}
\usepackage{xfrac}

\hypersetup{%
  pdflang={en-US},
  pdfkeywords={privacy; user study; exam proctoring},
  pdfdisplaydoctitle=true, %
  bookmarksnumbered,
  pdfborder = {0 0 0},
  colorlinks,
  linkcolor={black!80!black},
  citecolor={red!70!black},
  urlcolor={blue!70!black},
  pdfstartview={FitH},
  breaklinks=true,
  hypertexnames=false
}

\newcommand{\etal}{et~al.\xspace}

\newcolumntype{L}[1]{>{\hsize=#1\hsize\raggedright\arraybackslash}X}%
\newcolumntype{R}[1]{>{\hsize=#1\hsize\raggedleft\arraybackslash}X}%
\newcolumntype{C}[1]{>{\hsize=#1\hsize\centering\arraybackslash}X}%

\newcolumntype{H}{>{}@{}}

\newlist{compactitem}{itemize}{5}
\setlist[compactitem]{leftmargin=*, nosep}
\setlist[compactitem, 1]{label=\textbullet}
\setlist[compactitem, 2]{label=\textendash}
\setlist[compactitem, 3]{label=\textasteriskcentered}
\setlist[compactitem, 4]{label=\textperiodcentered}

\makeatletter
\let\orgItem\item
\NewDocumentCommand\fixedItem{ o }{%
   \IfNoValueTF{#1}%
      {\orgItem}%
      {\orgItem[#1]\def\@currentlabel{#1}}%
}
\makeatother

\newlist{questions}{enumerate}{3}
\setlist[questions]{align=left, labelwidth=2em, labelsep=.5em, listparindent=0pt, itemindent=0pt, leftmargin=!, before=\let\item\fixedItem}
\setlist[questions, 1]{labelindent=0pt, label=\textbf{Q\arabic*}, widest=99}
\setlist[questions, 2]{labelindent=-2.5em, label*=\textbf{\_\Alph*}, widest=26}
\setlist[questions, 3]{labelindent=-2.5em, label*=\textbf{\_\roman*}, widest=9}

\newlist{answers}{itemize}{1}
\setlist[answers]{leftmargin=*, nosep, align=left, label=$\bigcirc$}
\newlist{answers*}{itemize*}{1}
\setlist[answers*]{label=$\bigcirc$}

\microtypecontext{spacing=nonfrench}

\usepackage{titlesec}
\titlespacing{\paragraph}{0pt}{2ex}{1.5ex}

\titlespacing{\subparagraph}{0pt}{1.9pt}{1.5ex}
\titleformat*{\subparagraph}{\normalsize\em}
\begin{document}

\date{}

\title{\Large Educators’ Perspectives of Using (or Not Using) Online Exam Proctoring \thanks{This is an extended version of a paper appearing at the 32nd USENIX Security Symposium in August, 2023. Please reference: \smallskip \newline 
    David G. Balash, Rahel A. Fainchtein, Elena Korkes, Miles Grant, Micah Sherr, and Adam J. Aviv. ``Educators’ Perspectives of Using (or Not Using) Online Exam Proctoring.'' In the proceedings of the 32nd USENIX Security Symposium. USENIX Sec'. Aug.2023.}}

\author{
{\rm David G. Balash, Elena Korkes, Miles Grant, and Adam J. Aviv}\\
The George Washington University
\and
{\rm Rahel A. Fainchtein and Micah Sherr}\\
Georgetown University
} %

\def\plainauthor{Anonymous Submission}

\maketitle


\newcommand{\orgUnitTable}[0]{
\begin{table}[t]
\centering
\caption{The number of educators in each of the organizational units and the number of those educators who used online exam proctoring tools.}
\label{tab:org-unit-table}
\footnotesize
\begin{tabular}{l r r}
\toprule
  \multicolumn{1}{l}{\textbf{Organizational Unit}} &
  \multicolumn{1}{c}{\textbf{Educators}} &
  \multicolumn{1}{c}{\textbf{Used Tools}}  \\
\hline
\rowcolor[HTML]{EFEFEF}
Arts \& Sciences                            & 49   &   9  \\
Public Health                              & 16   &   0  \\
\rowcolor[HTML]{EFEFEF}
Medicine                  & 12   &   7  \\
Business                                   & 11   &   3  \\
\rowcolor[HTML]{EFEFEF}
International Affairs                      & 8    &   0  \\
Engineering \& Applied Sci.              & 8    &   1  \\
\rowcolor[HTML]{EFEFEF}
Nursing                                    & 6    &   5  \\
Education                                  & 6    &   0  \\
\rowcolor[HTML]{EFEFEF}
Professional Studies                      & 3    &   0  \\
Other                                     & 3    &   1  \\
\rowcolor[HTML]{EFEFEF}
Political Management              	       & 1    &   0  \\
Public Affairs                    & 1    &   0  \\
\rowcolor[HTML]{EFEFEF}
Arts \& Design                             & 1    &   0  \\
\hline
\textbf{Total}                                     & 125  &  26  \\
\bottomrule
\end{tabular}
\end{table}
}

\newcommand{\orgUnitTableUsedAndConsidered}[0]{
\begin{table}[t]
\centering
\caption{The number of educators in each of the organizational units~(\ref{appendix:main-survey:Q2}), the number of those educators who used online exam proctoring tools~(\ref{appendix:main-survey:Q6}), and the number of those educators who considered using the tools~(\ref{appendix:main-survey:N1}).}
\label{tab:org-unit-table}
\footnotesize
\begin{tabular}{l r r r}
\toprule
  \multicolumn{1}{l}{\textbf{Organizational Unit}} &
  \multicolumn{1}{c}{\textbf{Educators}} &
  \multicolumn{1}{c}{\textbf{Used}} &
  \multicolumn{1}{c}{\textbf{Considered}} \\
\hline
\rowcolor[HTML]{EFEFEF}
Arts \& Sciences                           & 49   &   9  & 14 \\
Public Health                              & 16   &   0  &  3 \\
\rowcolor[HTML]{EFEFEF}
Medicine                                   & 12   &   7  &  1 \\
Business                                   & 11   &   3  &  2 \\
\rowcolor[HTML]{EFEFEF}
International Affairs                      & 8    &   0  &  1 \\
Engineering \& Applied Sci.                & 8    &   1  &  2 \\
\rowcolor[HTML]{EFEFEF}
Nursing                                    & 6    &   5  &  1 \\
Education                                  & 6    &   0  &  0 \\
\rowcolor[HTML]{EFEFEF}
Professional Studies                       & 3    &   0  &  0 \\
Other                                      & 3    &   1  &  0 \\
\rowcolor[HTML]{EFEFEF}
Political Management              	       & 1    &   0  &  0 \\
Public Affairs                             & 1    &   0  &  1 \\
\rowcolor[HTML]{EFEFEF}
Arts \& Design                             & 1    &   0  &  0 \\
\hline
\textbf{Total}                             & 125  &  26  & 25 \\ 
\bottomrule
\end{tabular}
\end{table}
}

\newcommand{\orgUnitTableExtended}[0]{

\begin{table*}[htbp]
\centering
\caption{The number of educators in each of the organizational units and the number of those educators who used online exam proctoring tools and which respective tools they used.}
\label{tab:org-unit-table}
\footnotesize
\begin{tabular}{l r r r r r r r}
\toprule
  \multicolumn{1}{l}{\textbf{Organizational Unit}}  &
  \multicolumn{1}{c}{\textbf{Educators}}            &
  \multicolumn{1}{c}{\textbf{Used Tools}}           &
  \multicolumn{1}{c}{\textbf{Examsoft}}             &
  \multicolumn{1}{c}{\textbf{ProctorU}}             &
  \multicolumn{1}{c}{\textbf{Respondus}}            &
  \multicolumn{1}{c}{\textbf{RPNow}}                &
  \multicolumn{1}{c}{\textbf{Other}}                \\
\hline
\rowcolor[HTML]{EFEFEF}
Arts \& Sciences                           & 49   &   9   & 0 & 0 & 7 & 0 & 2  \\
Public Health                              & 16   &   0   & 0 & 0 & 0 & 0 & 0  \\
\rowcolor[HTML]{EFEFEF}
Medicine                                   & 12   &   7   & 0 & 0 & 5 & 2 & 0 \\
Business                                   & 11   &   3   & 0 & 0 & 3 & 0 & 0  \\
\rowcolor[HTML]{EFEFEF}
International Affairs                      & 8    &   0   & 0 & 0 & 0 & 0 & 0  \\
Engineering \& Applied Sci.                & 8    &   1   & 0 & 0 & 0 & 1 & 0  \\
\rowcolor[HTML]{EFEFEF}
Nursing                                    & 6    &   5   & 2 & 1 & 0 & 0 & 2 \\
Education                                  & 6    &   0   & 0 & 0 & 0 & 0 & 0  \\
\rowcolor[HTML]{EFEFEF}
Professional Studies                       & 3    &   0   & 0 & 0 & 0 & 0 & 0  \\
Other                                      & 3    &   1   & 0 & 0 & 0 & 0 & 1  \\
\rowcolor[HTML]{EFEFEF}
Political Management              	       & 1    &   0   & 0 & 0 & 0 & 0 & 0  \\
Public Affairs                             & 1    &   0   & 0 & 0 & 0 & 0 & 0  \\
\rowcolor[HTML]{EFEFEF}
Arts \& Design                             & 1    &   0   & 0 & 0 & 0 & 0 & 0  \\
\hline
\textbf{Total}                             & 125  &  26   & 2 & 1 & 15 & 3 & 5 \\
\bottomrule
\end{tabular}
\end{table*}
}

\newcommand{\subjectTable}[0]{
\begin{table}[ht]
\centering
\caption{The number of educators in each subject and the number of those educators who used online exam proctoring tools.}
\label{tab:subject-table}
\footnotesize
\begin{tabular}{l r r}
\toprule
  \multicolumn{1}{l}{\textbf{Subject}} &
  \multicolumn{1}{c}{\textbf{Educators}} &
  \multicolumn{1}{c}{\textbf{Used Tools}}  \\
\hline
\rowcolor[HTML]{EFEFEF}
S.T.E.M.                            & 40   &   5  \\
Medicine \& Health                  & 30   &   8  \\
\rowcolor[HTML]{EFEFEF}
Business                            & 14   &   6  \\
Government                          & 12   &   2  \\
\rowcolor[HTML]{EFEFEF}
Did not disclose                    & 7    &   3  \\
Arts                                & 5    &   0  \\
\rowcolor[HTML]{EFEFEF}
History                             & 5    &   0  \\
Languages                           & 4    &   0  \\
\rowcolor[HTML]{EFEFEF}
Communications                      & 4    &   1  \\
Gender Studies                      & 1    &   0  \\
\rowcolor[HTML]{EFEFEF}
Law             	                & 1    &   1  \\
Naval Science                       & 1    &   0  \\
\rowcolor[HTML]{EFEFEF}
Teaching                            & 1    &   0  \\
\hline
\textbf{Total}                      & 125  &  26  \\
\bottomrule
\end{tabular}
\end{table}
}

\newcommand{\subjectTableUsedAndConsidered}[0]{
\begin{table}[ht]
\centering
\caption{The number of educators in each subject~(\ref{appendix:main-survey:Q3}), the number of those educators who used online exam proctoring tools~(\ref{appendix:main-survey:Q6}), and the number of those educators who considered using the tools~(\ref{appendix:main-survey:N1}).}
\label{tab:subject-table}
\footnotesize
\begin{tabular}{l r r r}
\toprule
  \multicolumn{1}{l}{\textbf{Subject}} &
  \multicolumn{1}{c}{\textbf{Educators}} &
  \multicolumn{1}{c}{\textbf{Used}} &
  \multicolumn{1}{c}{\textbf{Considered}} \\
\hline
\rowcolor[HTML]{EFEFEF}
S.T.E.M.                            & 40   &   5    & 12 \\
Medicine \& Health                  & 30   &   8    &  5 \\
\rowcolor[HTML]{EFEFEF}
Business                            & 14   &   6    &  2 \\
Government                          & 12   &   2    &  2 \\
\rowcolor[HTML]{EFEFEF}
Did not disclose                    & 7    &   3    &  0 \\
Arts                                & 5    &   0    &  1 \\
\rowcolor[HTML]{EFEFEF}
History                             & 5    &   0    &  0 \\
Languages                           & 4    &   0    &  1 \\
\rowcolor[HTML]{EFEFEF}
Communications                      & 4    &   1    &  1 \\
Gender Studies                      & 1    &   0    &  1 \\
\rowcolor[HTML]{EFEFEF}
Law             	                & 1    &   1    &  0 \\
Naval Science                       & 1    &   0    &  0 \\
\rowcolor[HTML]{EFEFEF}
Teaching                            & 1    &   0    &  0 \\
\hline
\textbf{Total}                      & 125  &  26    & 25 \\
\bottomrule
\end{tabular}
\end{table}
}

\newcommand{\comfortTable}[0]{
\begin{table*}[ht]
\centering
\caption{Educators who reported using online exam proctoring tools ($n = 26$ of $125$) were asked to select how comfortable they would feel about using each monitoring type to monitor students during online proctored exams in their course~(\ref{appendix:main-survey:Q33}).  Most educators were comfortable with a lockdown browser. A live proctor not visible to students had the largest number of uncomfortable educators, followed by eye movement tracking and web browser history monitoring. The number of educators who reported enabling the monitoring type in~(\ref{appendix:main-survey:Q29}) are provided in parentheses.}
\label{tab:comfort-table}
\small

\begin{tabular}{llllll}
\toprule
\textbf{Monitoring Type}     & \textbf{Extremely} & \textbf{Somewhat} & \textbf{Neither comfortable} & \textbf{Somewhat} & \textbf{Extremely} \\
     & \textbf{Uncomfortable} & \textbf{Uncomfortable} & \textbf{nor uncomfortable} & \textbf{Comfortable} & \textbf{Comfortable} \\
\hline
\rowcolor[HTML]{EFEFEF}Live Proctor Not Visible     & 3 (0)                            & 5 (0)                           & 8 (0)                                          & 7 (2)                         & 3 (1)                          \\
Eye Movement Tracking        & 3 (0)                            & 5 (1)                           & 8 (1)                                          & 5 (3)                         & 5 (3)                          \\
\rowcolor[HTML]{EFEFEF}Web Browser History          & 3 (0)                            & 5 (0)                           & 4 (0)                                          & 5 (0)                         & 9 (3)                          \\
Webcam Recording             & 3 (0)                            & 3 (2)                           & 7 (3)                                          & 4 (3)                         & 9 (6)                          \\
\rowcolor[HTML]{EFEFEF}Screen Recording             & 2 (0)                            & 6 (1)                           & 6 (1)                                          & 4 (0)                         & 8 (6)                          \\
Microphone Recording         & 2 (0)                            & 6 (2)                           & 5 (1)                                          & 4 (2)                         & 9 (6)                          \\
\rowcolor[HTML]{EFEFEF}Face Detection               & 1 (0)                            & 7 (2)                           & 6 (1)                                          & 5 (3)                         & 7 (5)                          \\
Live Proctor Visible         & 1 (0)                            & 4 (0)                           & 8 (0)                                          & 7 (2)                         & 6 (4)                          \\
\rowcolor[HTML]{EFEFEF}Mouse Movement Tracking      & 0 (0)                            & 6 (0)                           & 11 (0)                                         & 6 (0)                         & 3 (1)                          \\
Keyboard Restrictions        & 0 (0)                            & 6 (0)                           & 6 (0)                                          & 6 (1)                         & 8 (2)                          \\
\rowcolor[HTML]{EFEFEF}Internet Activity Monitoring & 0 (0)                            & 4 (0)                           & 5 (0)                                          & 6 (1)                         & 11 (4)                         \\
Lockdown Browser             & 0 (0)                            & 1 (0)                           & 3 (2)                                          & 2 (2)                         & 20 (17)  \\                  
\bottomrule
\end{tabular}
\end{table*}
}

\newcommand{\likelySpringFallRegression}[0]{
\begin{table*}[htbp]
\centering
\caption[Likely Fall 2020 Spring 2021 regression]{\label{table:likely-fallspring-regression}
Binomial logistic regression model to describe which factors influenced the likelihood of using online exam proctoring tools for assessments assuming similar circumstances to those of the 2020/2021 academic year (\emph{Extremely likely} or \emph{Somewhat likely} responses to question \ref{appendix:main-survey:Q13}).  
The Aldrich-Nelson pseudo $R^2$ of the model is 0.65.
}
\small
\renewcommand{\arraystretch}{0.6}
\begin{tabular*}{\textwidth}{
l
@{\extracolsep{\fill}}
r
r
r
r
r
@{\extracolsep{6pt}}
l
}
  \toprule
{\textbf{Factor}} & {\textbf{Estimate}} & {\textbf{Odds ratio}} & {\textbf{Error}} & {\textbf{z value}} & {\textbf{Pr(\textgreater\textbar z\textbar)}} & {\textbf{ }} \\ 
  \midrule
  (Intercept)                                                               & -1.31 & 0.27    & 0.99 & -1.33  & 0.19    &   \\ 
  $\text{Live Proctor Visible} \in \{\textit{Ext Comf., Some Comf.}\}$      & -2.21 & 0.11    & 1.90 & -1.16  & 0.25    &   \\ 
  $\text{Live Proctor Not Visible} \in \{\textit{Ext Comf., Some Comf.}\}$  & 4.21  & 67.20     & 2.12 & 1.99   & 0.05    & * \\ 
  $\text{Internet Monitoring} \in \{\textit{Ext Comf., Some. Comf.}\}$      & 3.77  & 43.20     & 2.24 & 1.68   & 0.09    & . \\ 
  $\text{Webcam Recording} \in \{\textit{Ext Comf., Some. Comf.}\}$         & -3.48 & 0.03    & 2.24 & -1.56  & 0.12    &   \\ 
  $\text{Student Street Address} \in \{\textit{Slight Conc., Not Conc.}\}$  & 1.63  & 5.10     & 1.54 & 1.06   & 0.29    &   \\ 
  $\text{Less Cheating} \in \{\textit{Str Agree, Some Agree}\}$             & 2.41  & 11.20     & 1.37 & 1.76   & 0.08    & . \\ 
   \bottomrule
\end{tabular*}
\textbf{Signif. codes:} $\text{\enquote*{***}} \widehat{=} < 0.001$; $\text{\enquote*{**} } \widehat{=} <0.01$;$\text{\enquote*{*} } \widehat{=} <0.05$; $\text{\enquote*{.} } \widehat{=} <0.1$
\end{table*}
}

\newcommand{\likelyFallRegression}[0]{
\begin{table*}[htbp]
\centering
\caption[Likely fall 2021 return regression]{\label{table:likely-fall-2021-regression}
Binomial logistic regression model to describe which factors influenced the likelihood of using online exam proctoring tools for assessments assuming a similar teaching environment to Fall 2021 (\emph{Extremely likely} or \emph{Somewhat likely} responses to question \ref{appendix:main-survey:Q15}).
The Aldrich-Nelson pseudo $R^2$ of the model is 0.74.
}
\small
\renewcommand{\arraystretch}{0.6}
\begin{tabular*}{\textwidth}{
l
@{\extracolsep{\fill}}
r
r
r
r
r
@{\extracolsep{6pt}}
l
}
  \toprule
{\textbf{Factor}} & {\textbf{Estimate}} & {\textbf{Odds ratio}} & {\textbf{Error}} & {\textbf{z value}} & {\textbf{Pr(\textgreater\textbar z\textbar)}} & {\textbf{ }} \\ 
  \midrule
(Intercept)                                                                   & -3.09 & 0.01    & 1.68 & -1.84 & 0.07 & . \\ 
  $\text{Live Proctor Visible} \in \{\textit{Ext Comf., Some Comf.}\}$        & -1.90 & 0.15    & 1.67 & -1.13 & 0.26 &   \\ 
  $\text{Live Proctor Not Visible} \in \{\textit{Ext Comf., Some Comf.}\}$    & -3.17 & <0.01    & 2.42 & -1.31 & 0.19 &   \\ 
  $\text{Web Brwoser History} \in \{\textit{Ext Comf., Some Comf.}\}$         & 2.71  & 15.00     & 1.57 & 1.73  & 0.08 & . \\ 
  $\text{Face Detection} \in \{\textit{Ext Comf., Some Comf.}\}$              & 1.69  & 5.40     & 2.71 & 0.62  & 0.53 &   \\ 
  $\text{Mouse Movement Tracking} \in \{\textit{Ext Comf., Some Comf.}\}$     & -4.54 & <0.01    & 2.91 & -1.56 & 0.12 &   \\ 
  $\text{Student Room Scan} \in \{\textit{Slight Conc., Not Conc.}\}$         & 5.31  & 203.00      & 2.44 & 2.18  & 0.03 & * \\ 
   \bottomrule
\end{tabular*}
\textbf{Signif. codes:} $\text{\enquote*{***}} \widehat{=} < 0.001$; $\text{\enquote*{**} } \widehat{=} <0.01$;$\text{\enquote*{*} } \widehat{=} <0.05$; $\text{\enquote*{.} } \widehat{=} <0.1$
\end{table*}
}

\newcommand{\likelyFullReturnRegression}[0]{
\begin{table*}[htbp]
\centering
\caption[Likely full return regression]{\label{table:likely-full-return-regression}
Binomial logistic regression model to describe which factors influenced the likelihood of using online exam proctoring tools for assessments assuming the end of the pandemic and a full return to in person learning (\emph{Extremely likely} or \emph{Somewhat likely} responses to question \ref{appendix:main-survey:Q17}).
The Aldrich-Nelson pseudo $R^2$ of the model is 0.72.
}
\small
\renewcommand{\arraystretch}{0.6}
\begin{tabular*}{\textwidth}{
l
@{\extracolsep{\fill}}
r
r
r
r
r
@{\extracolsep{6pt}}
l
}
  \toprule
{\textbf{Factor}} & {\textbf{Estimate}} & {\textbf{Odds ratio}} & {\textbf{Error}} & {\textbf{z value}} & {\textbf{Pr(\textgreater\textbar z\textbar)}} & {\textbf{ }} \\ 
  \midrule
  (Intercept)                                                                  & -3.07  & 0.05    & 1.99 & -1.54   & 0.12 &   \\ 
  $\text{Live Proctor Visible} \in \{\textit{Ext Comf., Some Comf.}\}$         & -4.09  & 0.02    & 1.86 & -2.19   & 0.03 & * \\ 
  $\text{Eye Movement Tracking} \in \{\textit{Ext Comf., Some Comf.}\}$        & -1.84  & 0.16    & 2.60 & -0.70   & 0.48 &   \\ 
  $\text{Face Detection} \in \{\textit{Ext Comf., Some Comf.}\}$               & 6.32   & 557.00      & 3.25 & 1.95    & 0.05 & . \\ 
  $\text{Mouse Movement Tracking} \in \{\textit{Ext Comf., Some Comf.}\}$      & -6.00  & <0.01    & 3.29 & -1.82   & 0.07 & . \\ 
  $\text{Internet Monitoring} \in \{\textit{Ext Comf., Some. Comf.}\}$         & 4.47   & 87.00     & 2.71 & 1.65    & 0.10 & . \\ 
  $\text{Student Full Name} \in \{\textit{Slight Conc., Not Conc.}\}$          & 3.98   & 53.40     & 1.97 & 2.02    & 0.04 & * \\ 
  $\text{Student Photo ID} \in \{\textit{Slight Conc., Not Conc.}\}$           & -2.29  & 0.10    & 2.01 & -1.14   & 0.26 &   \\ 
   \bottomrule
\end{tabular*}
\textbf{Signif. codes:} $\text{\enquote*{***}} \widehat{=} < 0.001$; $\text{\enquote*{**} } \widehat{=} <0.01$;$\text{\enquote*{*} } \widehat{=} <0.05$; $\text{\enquote*{.} } \widehat{=} <0.1$
\end{table*}
}


\newcommand{\figMonitoringMethodsEnabled}[0]{
\begin{figure}[t]
\centering
\includegraphics[width=0.75\columnwidth]{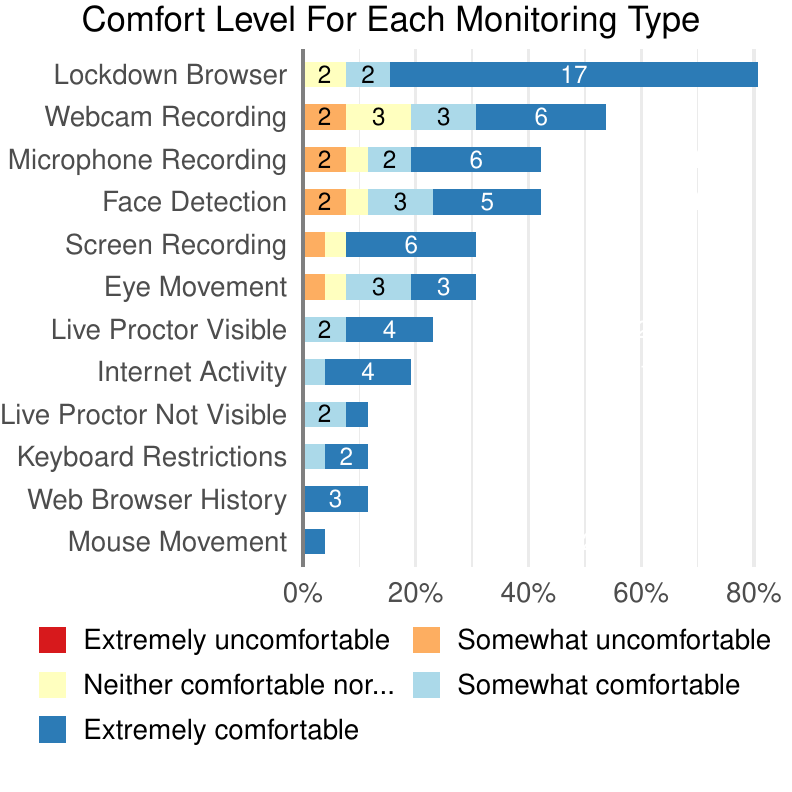}
\caption[Monitoring methods enabled bar plot]{\label{fig:monitoringTypesWithComfortMonitoringStudentsBar}
Educators who reported using online exam proctoring tools ($n = 26$ of $125$) were asked to select all monitoring methods they enabled~(\ref{appendix:main-survey:Q29}). Over 80\% of educators reported enabling the lockdown browser, and 50\% of educators enabled webcam recording during their online proctored exams. The educators were also asked to select how comfortable they would feel about using each monitoring type to monitor students during online proctored exams in their course~(\ref{appendix:main-survey:Q33}).  Most educators were comfortable with a lockdown browser. A live proctor not visible to students had the largest number of uncomfortable educators, followed by eye movement tracking and web browser history monitoring.
}
\end{figure}
}

\newcommand{\figComfortWithMonitoringStudents}[0]{
\begin{figure}[t]
\centering
\includegraphics[width=0.75\columnwidth]{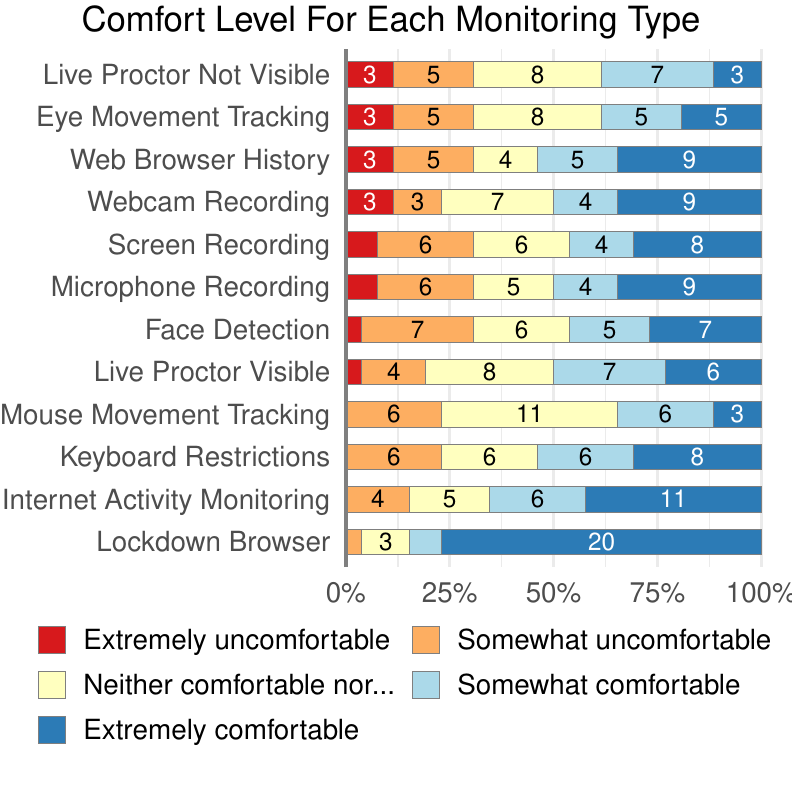}
\caption[Comfort with monitoring students bar plot]{\label{fig:comfortWithMonitoringStudentsBar}
Educators who reported using online exam proctoring tools ($n = 26$ of $125$) were asked to select how comfortable they would feel about using each monitoring type to monitor students during online proctored exams in their course~(\ref{appendix:main-survey:Q33}).  Most educators were comfortable with a lockdown browser. A live proctor not visible to students had the largest number of uncomfortable educators, followed by eye movement tracking and web browser history monitoring.
}
\end{figure}
}

\newcommand{\figConcernStudentsSharingInformation}[0]{
\begin{figure}[t]
\centering
\includegraphics[width=0.75\columnwidth]{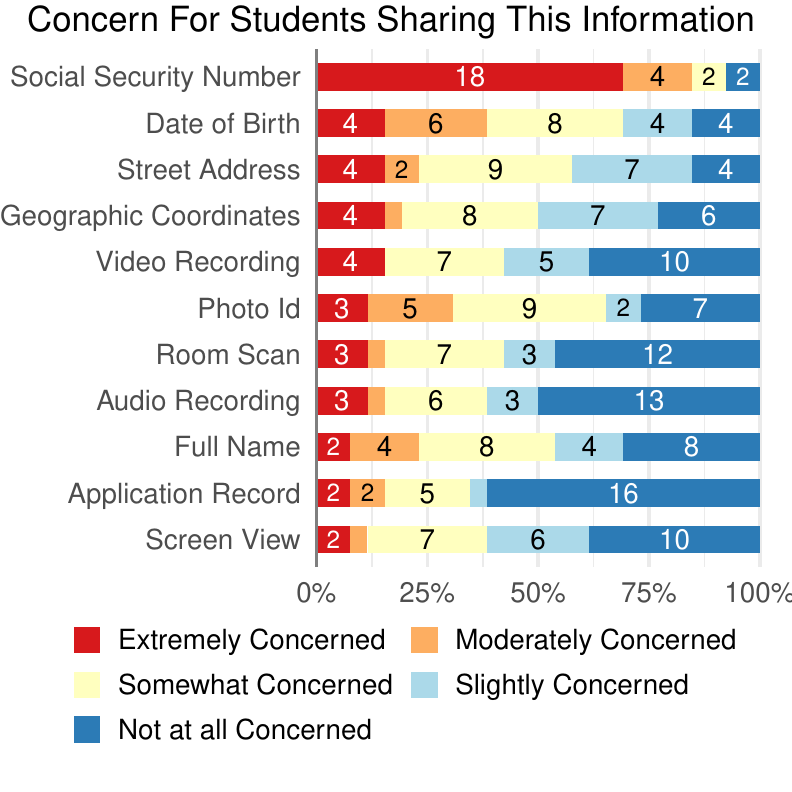}
\caption[Concern students sharing information bar plot]{\label{fig:concernStudentsSharingInformationBar}
Educators who reported using online exam proctoring tools ($n = 26$ of $125$) were asked to indicate how concerned they would be by students sharing each type of information with exam proctoring companies~(\ref{appendix:main-survey:Q34}). The largest number of educators ($n = 22$ of $26$;~85\%) were concerned with students sharing their social security number. While the fewest number of educators ($n = 3$ of $26$;~12\%) were concerned with students sharing their screen view.
}
\end{figure}
}

\newcommand{\figConcernStudentsSharingInformationExpanded}[0]{
\begin{figure}[t]
\centering
\includegraphics[width=\columnwidth]{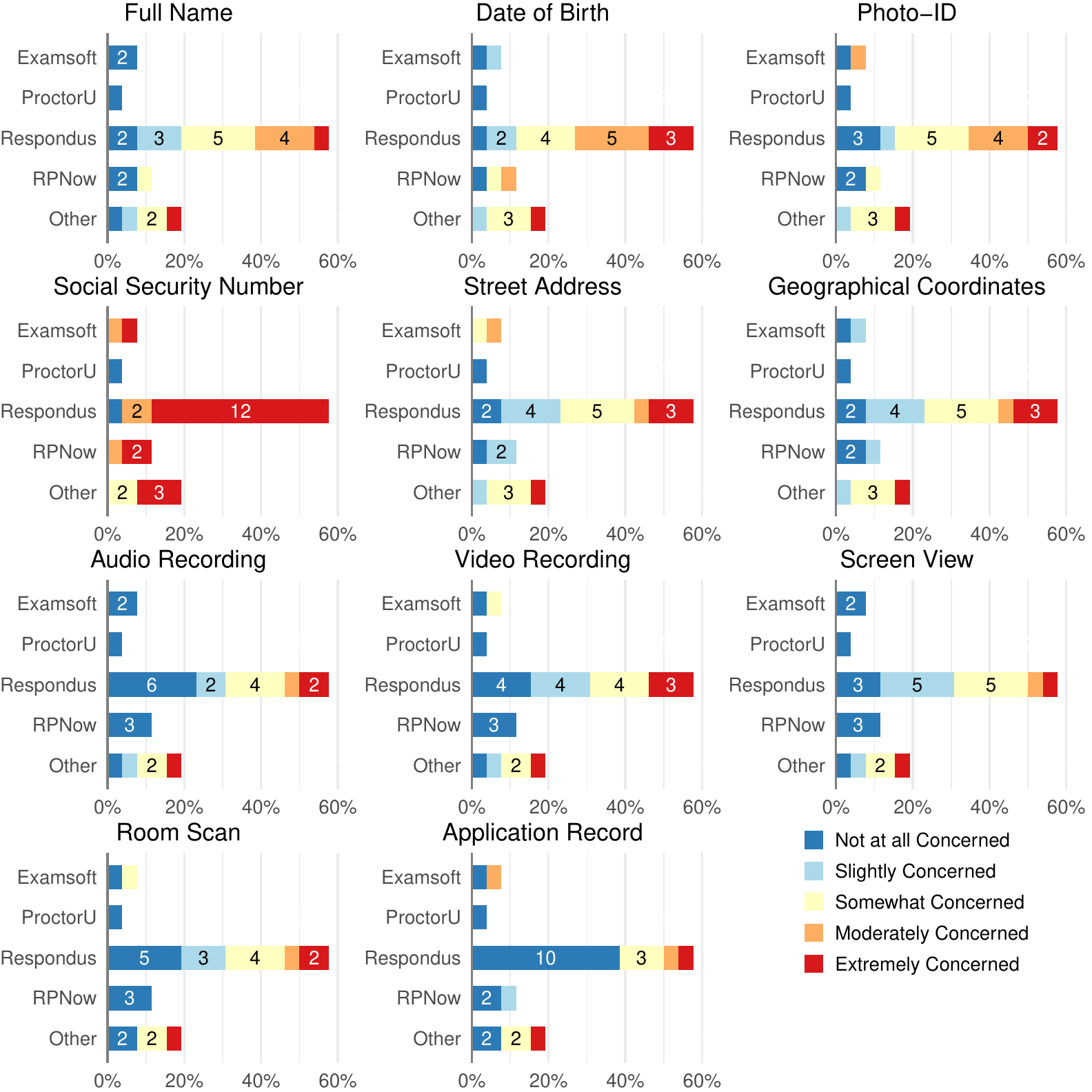}
\caption[Concern students sharing information bar plot]{\label{fig:concernStudentsSharingInformationBarExpanded}
Educators who reported using online exam proctoring tools ($n = 26$ of $125$) were asked to indicate how concerned they would be by students sharing each type of information with exam proctoring companies~(\ref{appendix:main-survey:Q34}). The concern for each type of information is shown by the specific exam proctoring tool that was used. The largest number of educators ($n = 22$ of $26$;~85\%) were concerned with students sharing their social security number. While the fewest number of educators ($n = 3$ of $26$;~12\%) were concerned with students sharing their screen view.
}
\end{figure}
}

\newcommand{\figAgreement}[0]{
\begin{figure*}[thbp]
\centering
\includegraphics{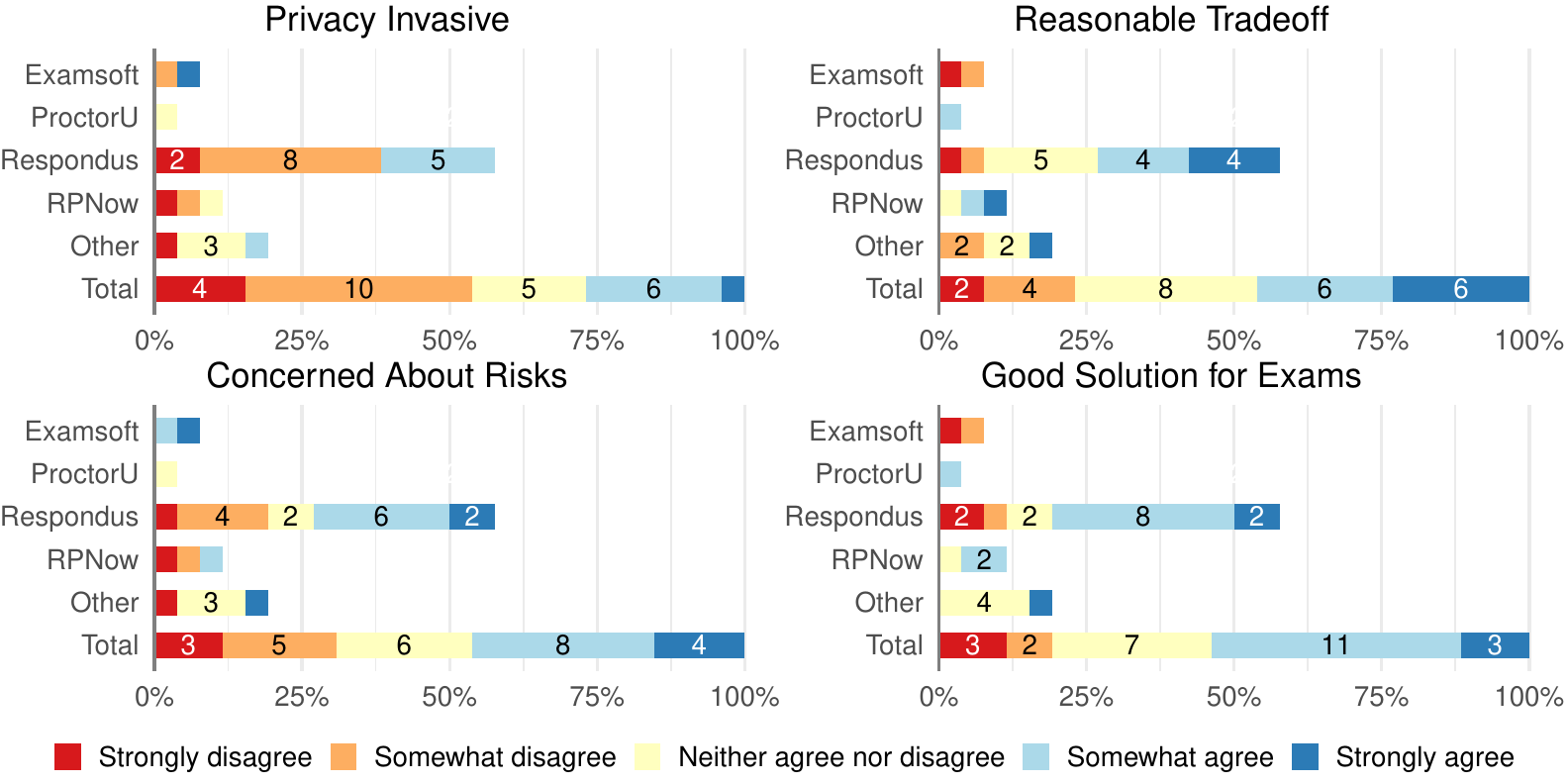}
\caption[Agree bar plot]{\label{fig:agreeBarChart}
Most of the ($n =26$) instructors who used online proctoring tools indicated they did not feel the proctoring tools they used were privacy invasive (\ref{appendix:main-survey:Q22}). However, as demonstrated by a plurality of respondents ($n = 12$;~46\%), instructors were (slightly) more concerned about the privacy risks the use of these tools posed to their students (\ref{appendix:main-survey:Q24}). Forth-six percent ($n = 12$) of instructors at least {\em somewhat agree} that the tools offered a reasonable tradeoff between student privacy and exam integrity (\ref{appendix:main-survey:Q23}). Most ($n = 14$;~54\%) participants felt the tools they used offered a good solution for remotely monitoring exams  (\ref{appendix:main-survey:Q25}).
} 
\end{figure*}
}

\newcommand{\figLikelyAlluvium}[0]{
\begin{figure*}[thbp]
\centering
\begin{subfigure}[t]{\columnwidth}
\centering
    \includegraphics[width=0.75\textwidth]{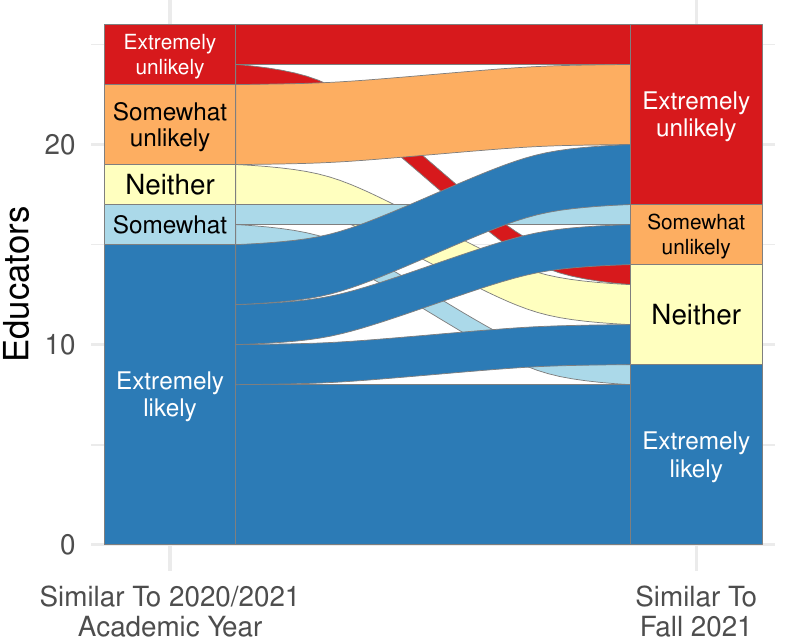}
    \caption[Likely to use exam proctoring 2020/2021 academic year to Fall 2021 alluvial plot]{\label{fig:likelyCircumstanceFullVirtualToHybridAlluvial}
        2020/2021 academic year compared with Fall 2021.
    }
\end{subfigure}
\hfill
\begin{subfigure}[t]{\columnwidth}
\centering
    \includegraphics[width=0.75\textwidth]{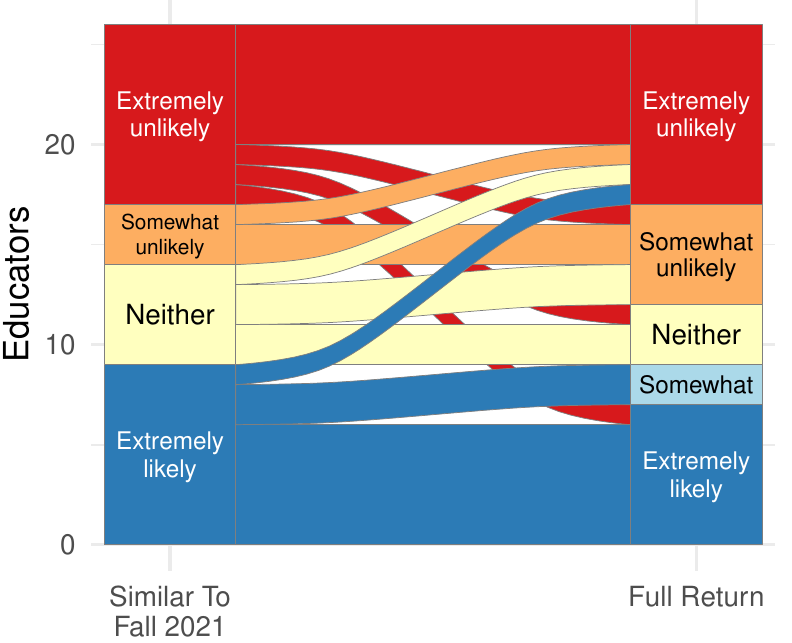}
    \caption[Likely to use exam proctoring fall 2021 to full in-person alluvial plot]{\label{fig:figLikelyCircumstanceHybridToFullInPersonLearning}
        Fall 2021 compared with full return to in-person learning.
    }
\end{subfigure}
\caption[Alluvium plots of likely to use exam proctoring tools]{\label{fig:likelyToUseExamProctoringToolsAlluvium}
Detailed visualization of how likely educators are to use remote proctoring under circumstances similar to 
\begin{enumerate*}[label=(\alph*)]
\item 2020/2021 academic year and Fall 2021 (\ref{appendix:main-survey:Q13} \& \ref{appendix:main-survey:Q15}) and 
\item Fall 2021 and a full return to in-person learning (\ref{appendix:main-survey:Q15} \& \ref{appendix:main-survey:Q17}).
\end{enumerate*}
}
\end{figure*}
}

\newcommand{\figCatchCheating}[0]{
\begin{figure}[t]
\centering
\includegraphics[width=0.75\columnwidth]{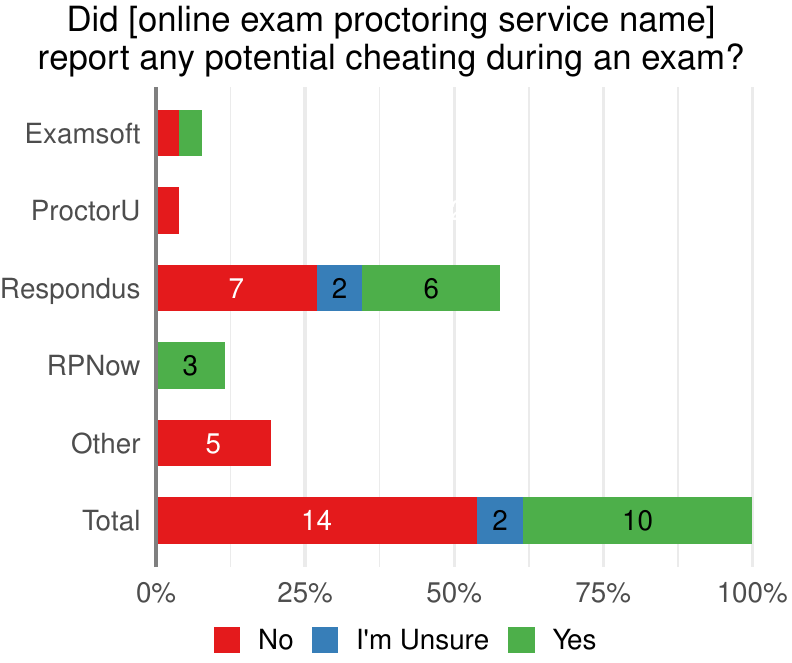}
\caption[Cheating will catch this percent. Report potential cheating.]{\label{fig:catchCheating}
When asked if the online exam proctoring tool they used reported any potential cheating \ref{appendix:main-survey:Q21}, 38\% ($n = 10$ of $26$) said that it had.}

\end{figure}
}

\newcommand{\figLessLikelyToCheat}[0]{
\begin{figure}[t]
\centering
\includegraphics[width=0.75\columnwidth]{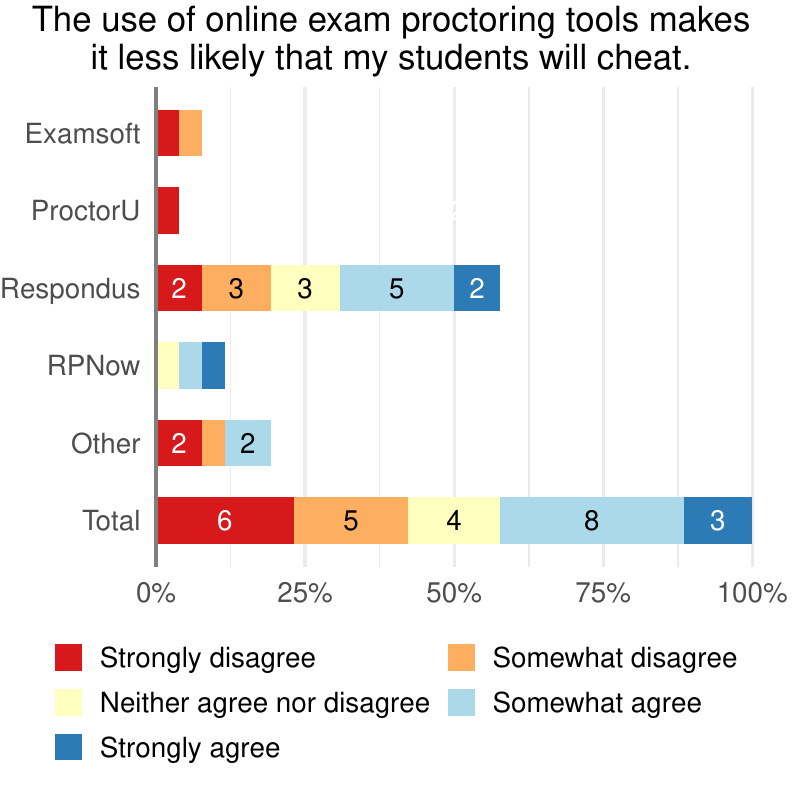}
\caption[Cheating will catch this percent. Report potential cheating.]{\label{fig:lessLikelyToCheat}
When asked if the use of the tools makes it less likely that students will cheat \ref{appendix:main-survey:Q19}, 42\% ($n = 11$ of $26$) at least {\em somewhat agree}.}

\end{figure}
}

\newcommand{\figSoftwareConcern}[0]{
\begin{figure}[t]
\centering
\includegraphics[width=0.75\columnwidth]{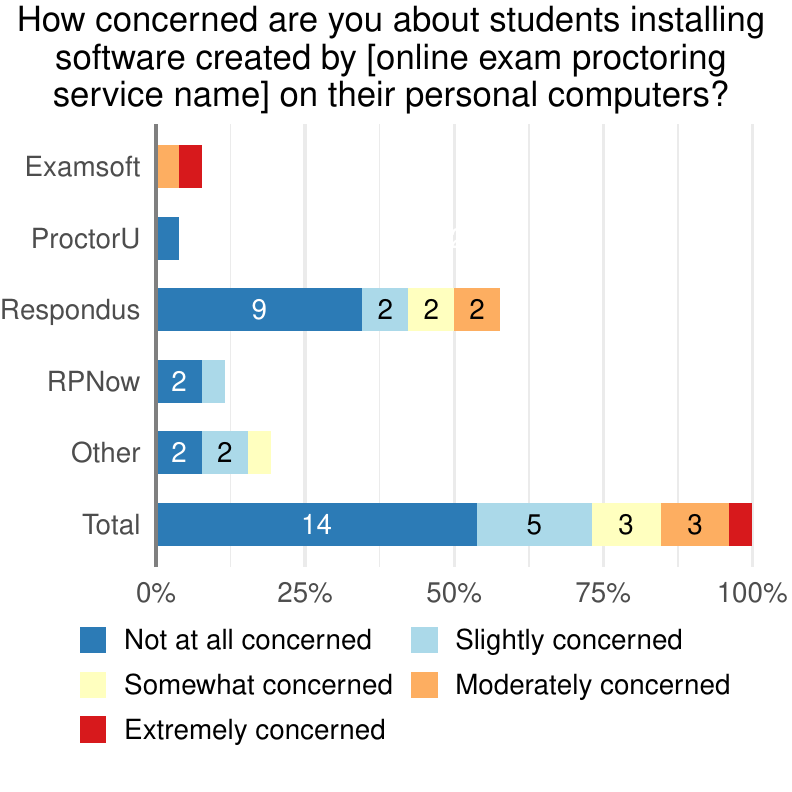}
\caption[Concern about students installing software.]{\label{fig:softwareConcern}
When educators who had used online exam proctoring were asked to report their level of concern about students installing software created by online exam proctoring services on their personal computers \ref{appendix:main-survey:Q27}, more than half ($n = 14$ of $26$;~54\%) reported that they were {\em not at all concerned}, while 27\% ($n = 7$ of $26$) were at least {\em somewhat concerned}.}
\end{figure}
}

\newcommand{\figPrivacyTradeoffs}[0]{
\begin{figure}[t]
\centering
\includegraphics[width=0.75\columnwidth]{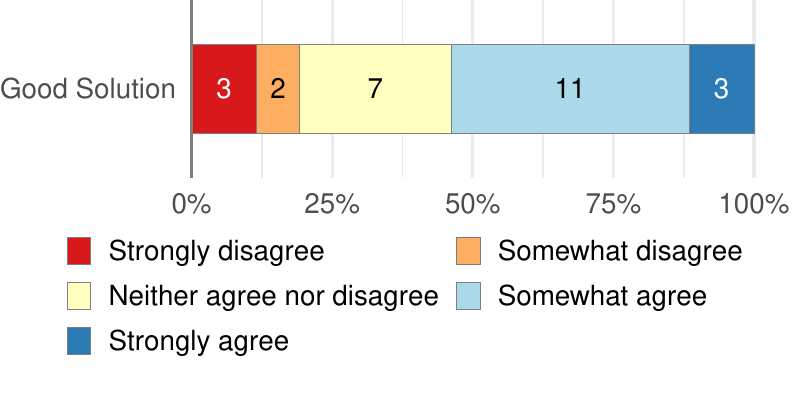}
\caption[Good Solution] {\label{fig:privacyTradeoffs}
Of the ($n = 26$) participants who used online proctoring, most ($n = 14$;~54\%) participants felt the (respective) tools they used offered a good solution for remotely monitoring exams  (\ref{appendix:main-survey:Q25}).

}
\end{figure}
}

\newcommand{\figPrivacyInvasive}[0]{
\begin{figure}[t]
\centering
\includegraphics[width=0.75\columnwidth]{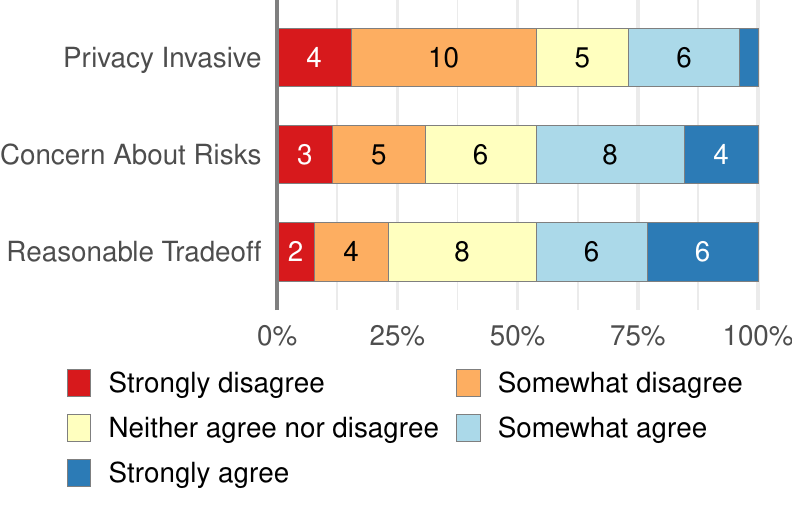}
\caption[Privacy Invasive, Concern About Risks.]{\label{fig:privacyInvasive}
Most of the ($n =26$) instructors who used online proctoring tools indicated they did not feel the (respective) proctoring tools they used were privacy invasive (\ref{appendix:main-survey:Q22}). However, as demonstrated by a plurality of respondents ($n = 12$;~46\%), instructors were (slightly) more concerned about the privacy risks the use of these tools posed to their students (\ref{appendix:main-survey:Q24}).
Forth-six percent ($n = 12$) of instructors at least {\em somewhat agree} that the tools offered a reasonable tradeoff between student privacy and exam integrity (\ref{appendix:main-survey:Q23}).  
}
\end{figure}
}
\begin{abstract}

The onset of the COVID-19 pandemic changed the landscape of education and led to increased usage of remote proctoring tools that are designed to monitor students when they take assessments outside the classroom. While prior work has explored students' privacy and security concerns regarding online proctoring tools, the perspective of educators is under explored. Notably, educators are the decision makers in the classrooms and choose which remote proctoring services and the level of observations they deem appropriate. To explore how educators balance the security and privacy of  their students with the requirements of remote exams, we sent survey requests to over 3,400 instructors at a large private university that taught online classes during the 2020/21 academic year. We had $n=125$ responses: 21\% of the educators surveyed used online exam proctoring services during the remote learning period, and of those, 35\% plan to continue using the tools even when there is a full return to in-person learning. Educators who use exam proctoring services are often comfortable with their monitoring capabilities. However, educators are concerned about students sharing certain types of information with exam proctoring companies, particularly when proctoring services collect identifiable information to  validate students' identities. Our results suggest that many educators developed alternative assessments that did not require online proctoring and that those who did use online proctoring services often considered the tradeoffs between the potential risks to student privacy and the utility or necessity of exam proctoring services.

\end{abstract}
\section{Introduction}

The initial surge of the COVID-19 pandemic upended education, leading many schools to quickly switch to remote teaching in the Spring of 2020~\cite{eduOnlineCovid}, and many universities and colleges maintained remote learning into the 2020/21 academic year. 
This massive migration to online learning environments led to a corresponding increase in the use of remote educational technologies. 

One such remote learning technology that saw a dramatic increase in use during remote instruction is online exam proctoring tools. 
Based on the analysis of the Chrome browser extension reviews, Balash \etal found explosive growth (720\%) of online proctoring beginning at the start of the COVID-19 pandemic~\cite{balash-21-exam}. 
This is in line with a poll by Grajek that found that 77\% of colleges and universities made use of or were planning to use online proctoring~\cite{educausepoll}.

 By design, remote proctoring systems are invasive.  Given their capabilities to monitor and limit functionality where installed, students and privacy advocates have raised concerns about their security and privacy properties.  As highlighted by the media coverage of remote proctoring tools, these concerns %
were not unfounded: 
Since the tools' widespread adoption at the beginning of the pandemic, reports uncovered major security and privacy incidents involving Proctorio and ProctorU, two widely used invigilation tools. These included a major data breach of ProctorU in which 444,000 users' personally identifying information was leaked online and a security vulnerability within Proctorio that allowed hackers to remotely activate the software on computers in which it was installed \cite{bleepingProctorUBreach2020,Proctorio2021ndl,chroniclesProctorio2022}.
More recently, Burgess \etal~\cite{burgess2022watching} disclose several security and privacy issues, including concerns about how remote proctoring systems use facial recognition.

In a survey of students who experienced remote proctoring, Balash \etal found that many students had both privacy and security concerns with the tools~\cite{balash-21-exam}. In particular, student participants often felt they had no choice but to use the tools or that they trusted these proctoring services because of their academic institutions' support for them. Despite some students' trust in these tools' security, Coheny~\cite{cohney-2021-harms} show evidence indicating that these  tools may not be as trustworthy as students suspect. Specifically, they find that many collaborative tools used in remote classrooms collect information about students that often does not align with educational expectations. However, the tools analyzed in their study focus on collaboration tools that were not designed for academic settings.

As such {\em educators'} perspectives of online exam proctoring services
remains unexplored.  Educators` perspectives are of particular importance given their roles in both the choice to use (or not use) an online proctoring tool and its associated monitoring. 
In this paper we seek to answer the following research questions about how educators consider privacy and security in the context of online proctoring services:

\begin{enumerate}[leftmargin=*,noitemsep,label=\textbf{RQ\arabic*}]
    \item\label{RQ:educator-perceptions} What are educators' perceptions of online proctoring services?
    \item\label{RQ:privacy-concerns} Do educators consider student privacy and security concerns when deciding to use (or not use) an online proctored exam and while setting up the exam proctoring parameters?
    \item\label{RQ:method-selection} Which proctoring methods do educators select to proctor their online exams?
\end{enumerate}

To answer these research questions, we executed a campus-wide recruitment of instructors at the George Washington University  who taught courses during the remote learning period of the 2020/21 academic year. This involved inviting 3,460 educators to participate in an IRB-approved survey, of which $n = 125$ participants responded with their justifications for using or not using online proctoring. The survey captured responses from  the university's 12 organizational units, senior and junior faculty, as well as graduate educators. %
Despite our small sample (approximately $\sfrac{1}{5}$ of respondents opted to use online proctoring), our results offer important and timely insights into how educators at a large-private university understood online proctoring tools and their motivations for using (or not using) these tools during a challenging period.

We found that a small but substantial number (21\%) of educators used exam proctoring tools during the 2020/21 academic year.
The most common reasons for using remote proctoring were to stop or deter cheating, to comply with COVID-19 safety protocols, to maintain exam integrity, and to be fair to students. 
In contrast, 79\% of respondents did not use online exam proctoring tools during this same period.
Many chose not to use remote proctoring tools due to their potential harms to students,  negative impacts on trust between students and educators, student privacy concerns, ineffectiveness against cheating, and the availability of alternative modes of assessment, such as open-book exams and projects.

Both educators who used online proctoring and those who did not reported privacy concerns with using exam proctoring services.  These concerns centered on webcam and audio recordings taken by a third party, intrusive monitoring measures, information sharing requirements, particularly those of personally identifying information for verification of student identities, and the invasion of student privacy  as students take these exams in their homes. 
Educators also expressed concerns about the security implications of students having to install exam proctoring software on their computers. Many highlighted the software's monitoring capabilities and its ability to disable system functionality. %

While many educators chose to modify their assessments rather than use online exam proctoring, some educators were required to use online exam proctoring tools. 
Specifically, these educators reported departmental mandates for the use of exam proctoring services in their courses, or requirements to administer the standardized tests in their field, such as nursing, that were only offered by testing companies that use online exam proctoring technologies.
This led to educators being forced to use these proctoring tools despite having reservations about their use and concerns about their impact on students.

\section{Related Work}\label{sec:background}
\label{sec:related-work}

Online proctoring tools have been the subject of heavy scrutiny from both the media~\cite{swauger2020,nyt1,nyt2,wapo1} and education researchers~\cite{woldeab201921st,Hbert2021OnlineRP,GenisGruber2022ChallengesOO,Meulmeester2021MedicalSP,Silverman2021WhatHW,Schultz2022PerilsAP}. Below, we frist review the literature on the whether proctoring is needed to ensure academic integrity and are efficacious in doing so, and how they impact students have been at the center of debates on their role within remote learning. Following, we will discuss more recent work studying the security and privacy impact of this technology.

\paragraph{Effectiveness in Academic Integrity}
When it comes to the role of online proctroing in online learning, Harton \etal find that despite vast improvements in remote learning tools, university instructors and students show a strong bias towards the beliefs that online courses are more conducive to academic dishonesty and that cheating occurs more often in online settings~\cite{harton2019}. However, studies comparing the rates of academic dishonesty in face to face courses to its prevalence in online courses have found mixed results:
Watson and Sottile find that while students more readily admit to academic dishonesty in face to face classes, they are more likely to cheat during online exams~\cite{watson2010cheating}. %
In contrast, Grijalva \etal\cite{grijalva2006} find no significant difference in rates of academic dishonesty in both course formats.

Gudi\~no Paredes \etal, who evaluated remote proctored exams' usage in graduate learning via a questionnaire-based study, find that the tools enforce academic integrity, but that students' honesty is neither driven by their moral compasses nor by their desire to learn\cite{gudinoparedes2021}. Instead, as Gudi\~no Paredes \etal explain, students lack opportunities to cheat due to constraints implemented within the tools and feel obliged to behave with integrity lest they be caught and punished by the software. Moreover, they find that these tools appear to negatively impact students' learning or motivation to learn and raise concerns about student privacy. They therefore recommend educators carefully consider their motivation to use remote proctoring before choosing to use these tools.

\paragraph{Student Performance Under Proctoring}
Several studies~\cite{goedl2020study,daffin2018comparing,seife2020,Davis2016RemotePT} have found that 
student performance was significantly better on unproctored remote exams than on remotely proctored ones.
Seife and Stockton~\cite{seife2020} further find that scores on remotely proctored exams are more closely correlated with predictive attributes of student performance, such as their ratings of human capital, which measures their general ability level.  This, they argue, shows evidence that academic misconduct is likely quite pervasive, and that this pays off handsomely for dishonest students.
Moreover, Wuthisatian\cite{wuthisation2020} finds that students generally performed better on in-person exams than they did when the same exam was taken with remote online proctoring. 

In contrast however, Hylton \etal do not find a significant difference in student performance between remote exams that use video monitoring and those that do not. Despite this, they note that students in non-video monitored exams took longer to complete their exams on average\cite{hylton2016utilizing}.  
Rios and Liu similarly find that student performance on low stakes exams is not impacted by the use of online proctoring. However, unlike Hylton \etal, they do not observe any differences in the amount of time students take to complete their exams\cite{rios2017}. 
In a study of the differences between testing centers and remote proctoring by Cherry \etal\cite{Cherry2021DoOF}, average scores achieved across the two proctoring modes were similar.

\paragraph{Privacy, Security and Ethics of Remote Proctoring}
 
 Online remote proctoring and other online learning technology have received considerable attention due to the pandemic.  Coghlan \etal\cite{coghlan2020good}, in their opinion piece, highlight proctoring tools' reliance on artificial intelligence and the ethical challenges this can raise. They explore an ethical framework for determining when and how to use these tools. In their opinion article, Swauger argues the underlying algorithms for remote monitoring and invigilation contain implicit negative biases %
 and that these tools unfairly penalize students who do not meet their biased baseline\cite{swauger2020}.

Despite these concerns, research on these tools' security and how users perceive their security
has been sparse. Balash \etal and Kharbat and Abu Daabes independently study student perceptions of the tools, and find a high prevalence of privacy concerns \cite{kharbat2021proctored, balash-21-exam}. Balash \etal, who also analyze the security and privacy of several tools, specifically focus on how students' security and privacy concerns compare with the security vulnerabilities they encounter\cite{balash-21-exam}.  Neither Balash \etal nor Kharbat and Abu Daabes consider the perceptions of educators, the focus of this paper.

Recent work by Burgess \etal~\cite{burgess2022watching}, performed a technical S\&P analysis of four proctoring suites used in high stakes law licensing exams, such as the Bar Exam and entrance exams. They identified numerous privacy and security risk, including around facial recognition. In this paper, we investigate the educator's perspective on eight general purpose exam proctoring software suites, which are non-overlapping with the suites studied by Burgess \etal. Such a technical investigation, though, would likely be fruitful of the common university remote exam proctoring products.

Similar to this paper is the study by Cohney \etal that also considers instructor and faculty perceptions of remote learning tools~\cite{cohney-2021-harms},  but not specifically remote proctoring tools. The focus of Cohney \etal is on the use of tools that were not originally designed for educational use, but were adopted hastily amidst the pandemic to accommodate the need for fully remote learning. This includes using Zoom, Google Drive, and other collaboration platforms whose privacy standards and data collection practices do not match the expectations of the classroom. In contrast, in this paper, we focus on instructors perceptions, and more specifically, on educators' perceptions of remote proctoring tools and their security and privacy attributes, both why they choose to use them and why not. Cohney \etal, in contrast, identifies such tools, but was not the primary focus of the study.

\section{Survey Methodology}\label{sec:method}

We conducted an online survey to evaluate university educator perceptions of online exam proctoring tools. 
Here we describe the survey's procedures, recruitment, limitations, and ethics. Survey results are presented in Section~\ref{sec:results}.

\paragraph{Study Procedure}

Below we outline the survey. The full text can be found in Appendix~\ref{appendix:main-survey}. 

\begin{enumerate}[nosep]

\item Informed Consent: 
The university educators were asked to consent to the study.  The consent included that participants would answer questions about their experience with online exam proctoring services. 

\item Eligibility Screening:
To be eligible to complete the survey, participants were required to assert that they were either full-time faculty or part-time adjunct faculty at the university.

\item Background:
The educators were then asked to optionally provide their associated organizational unit or school at the university, as well as the subject area(s) they taught during the 2020/21 academic year

\item Awareness of Technology:
The university educators were asked about their awareness of the online exam proctoring tools available at the university during the 2020/21 academic year and their understanding of how online exam proctoring tools work. Next participants were asked which specific online exam proctoring tools, if any, they used in administering assessments  during the 2020/21 academic year.

\item Use and Perceptions of Online Exam Proctoring Tools:
Educators were asked which proctoring services they most recently used, what factors they considered, and the type and number of assessments administered with online proctoring. Next the educators were asked about both the benefits and drawbacks of online exam proctoring and under what conditions they were likely to use online exam proctoring in the future. 

\item Proctoring Effectiveness:
We then asked the educators about the effectiveness of online exam proctoring tools at preventing and catching cheating on assessments. 

\item Review of Proctoring Tools Used:
Educators were asked questions to assess the specific exam proctoring tools they reported to have recently used during the 2020/21 academic year. This included questions about the educators' views of the privacy and security of the online exam proctoring software, its effectiveness, and the potential tradeoff between student privacy concerns and the integrity of the examination being administered. 

\item Online Exam Proctoring Methods:
In this part of the survey we investigate the methods used by online exam proctoring services to monitor student test takers.  Educators were asked which exam monitoring methods they enabled in their proctored exams, the effectiveness of these methods, their comfort using the methods, and if they would change methods for future exams. 

\item Privacy Concerns:
Finally, the survey concluded by asking educators about their concerns for their students' privacy when students are required to share information with exam proctoring companies.

\end{enumerate}

\paragraph{Recruitment}

We worked with the George Washington University's administration to approve and coordinate the survey, in addition to receiving IRB approval (NCR202908). In turn the university provided our research team with an email list of all instructors who had taught a course during the 2020/21 academic year, totally 3,460 individuals. Recruitment occurred over a fifteen day period starting December 1st 2021. We sent out 3,460 emails and had 152 educators respond to the study, a response rate of 4.4\%. Of the 152 educators who responded, 125 completed the study. Recruitment emails was sent by and the survey was hosted on the university's Qualtrics account. Participants who completed the survey were given the opportunity to enter a drawing for a \$50 USD Amazon gift card with a 1 in {20} chance of winning. On average, it took {16.3} minutes (SD=37.4) to complete the study.

Note that many of the instructors who were contacted may either no longer be at the university or did not teach classes involving exam or quiz assessments (e.g., instead using grading based solely on term-papers) that would render them eligible to complete the survey. These instructors likely self-selected out of the survey, and thus the true number of eligible participants and the true response rate to the survey is difficult to determine. However, we believe we captured a reasonably representative cross-section of the university's instructors during this period, both in terms of educators who chose to use online proctoring and those who chose not to use it. But it is also important to acknowledge that there are likely some perspectives that may be over- or underrepresented due to self-selection both to take and not take the survey.

\paragraph{Analysis Methods} 
When presenting quantitative results the analysis is provided in context. 
For qualitative responses, we conducted open coding to analyze 14 free-response questions. 
A primary coder from the research team crafted a codebook and identified descriptive themes by coding all responses to each question.
For the 10 open-ended questions answered by participants who did use exam proctoring tools, a secondary coder (also a member of the research team) coded all responses.
For the 4 open-ended questions responded to by educators who did not use proctoring tools, a secondary coder coded a 20\% sub-sample as a consistency check. In each case, the secondary coder provided feedback on the codebook, and inter-rater reliability was calculated on each round until Cohen's $\kappa \geq 0.7$. Overall, the mean Cohen's $\kappa=0.8$, indicating substantial agreement between coders. In the results presented below, we use the primary-coders application of the final codebook for any counts or themes presented.

\paragraph{Ethical Considerations}
The study protocol was approved by our Institutional Review Board (IRB) with approval number NCR202908, and all collected data is associated with random identifiers. For participants who wanted to enter the drawing for a chance to win the Amazon gift card we created an entry form that was separate and not linked to the survey.%
We also considered that some educators may not want to share how they managed student academic integrity for online classes or their specific academic department or subject area and so we made those questions optional.

\paragraph{Limitations}
Our study is limited in its recruitment, particularly instructors at a single academic institution in the U.S. While our study offers a unique perspective at our institution, which is a large private university, we cannot claim full generalizability of the results within or beyond  the institution as it is difficult to know the true number of eligible participants who considered or used online proctoring during the 2020/21 academic year. Despite this limitation, we believe that these results offer new insights and are likely representative of common attitudes and themes among educators about online proctoring and their choices to use or not use these products. However, we cannot conclude that the these themes occur at the same proportions beyond our sample. We attempt to note this limitation throughout when discussing proportional results. This is particularly true for those instructors who indicate that they do use online proctoring, and we cannot be confident that quantitative results, e.g., Likert responses, will be consistent in a larger sample. Although qualitative themes likely express dominant views in this subgroup, we may not capture all minority themes. Throughout the following section we acknowledge these limitations.

There is also limitations with the size of our recruitment. We got approval to send recruitment emails to \textit{all} instructors during the online-instruction period at our institution, which included 3,460 individuals. Even within this pool, finding college level educators that use exam proctoring turns out to be a difficult-to-reach population. As with any online survey without direct recruitment, response rates can be small, and then within those responses, we are further seeking a set of educators who used online proctoring.  We acknowledge that the result is a sample of educators who actually used proctoring is smaller than desirable, but when targeting hard-to-reach populations (namely, college educators that use exam proctoring), exploratory studies like this one, even with smaller samples, provide important and relevant themes.

Importantly, the goal of this study is not only on the educators who did use proctoring, but also those that choose not to and their security and privacy reasons for that choice. We were able to recruit 99 participants who decided against using online exam proctoring tools to  provide important insight into their decision making processs.

Finally, we are limited by the fact that this study relies on self-reported behavior. We cannot verify that the participants actually used remote proctoring tools to proctor an online exam or which monitoring methods they enabled. Finally, responses can suffer from social desirability and response bias, leading participants to over describe their awareness of online exam proctoring as they may believe that this is the expectation of the researchers. Such biases may be most present when participants indicate concerns.

\section{Results}\label{sec:results}

All of the educators in our study taught a course during the 2020/21 academic year at a the George Washington University.
Educators from twelve of the university's organizational units or schools were represented (see \autoref{tab:org-unit-table}), with the largest percentages from the College of Arts \& Sciences ($n = {49}$;~39\%), School of Public Health ($n = {16}$;~13\%), School of Medicine ($n = {12}$;~10\%), and the School of Business ($n = {11}$;~9\%) (\ref{appendix:main-survey:Q2}).

During the 2020/21 academic year the educators surveyed taught a wide range of subjects.
The most common of which were science, technology, engineering, and mathematics ($n = {40};~32\%$), health ($n = {30};~24\%$), business ($n = {14};~11\%$), and government ($n = {12};~10\%$)  (\ref{appendix:main-survey:Q3}).  
For the full results see \autoref{tab:subject-table}.
Twenty-one percent ($n = {26}$) of educators who responded to our survey used online exam proctoring tools to assist in administering assessments during the 2020/21 academic year (\ref{appendix:main-survey:Q6}).

At the time of the study, eight online exam proctoring tools were available at the university. 
Educators reported being most aware of exam software by Respondus ($n = {56};~45\%$), ProctorU ($n = {16};~13\%$), and Examsoft ($n = {9}; ~7\%$) (\ref{appendix:main-survey:Q4}).
Of the educators who reported using exam proctoring software, the largest number reported using Respondus ($n = {15}$ of ${26};~58\%$), followed by RPNow ($n = {3}$ of ${26};~12\%$), Examsoft ($n = {2}$ of ${26};~8\%$), and Proctorio ($n = {2}$ of ${26}; ~8\%$), for their most recent proctored online exam (\ref{appendix:main-survey:Q7}). 
Most ($n = {23}$ of ${26};~88\%$) educators who used online exam proctoring tools used them for administering course exams (e.g. test, midterm exam, final exam) (\ref{appendix:main-survey:Q9}). 
Among educators who used online proctoring sixty-five percent reported ($n = {17}$ of ${26}$) having administered five or more online proctored assessments  (\ref{appendix:main-survey:Q10}).

\orgUnitTableUsedAndConsidered
\subjectTableUsedAndConsidered

\subsection{RQ1: Educators' Perceptions}
\label{sec:rq1}

\paragraph{Educator Understanding of Exam Proctoring Tools}
We asked educators to describe in their own words how online proctoring tools work~(\ref{appendix:main-survey:Q5}). 
Many ($n = {65}$) educators described the ways the proctoring tools monitor a student's activity and behavior during an exam. 
Educator P2 (College of Arts \& Sciences) responded, ``Monitor student's actions and movements (and room content) to make sure they are not cheating on an exam.'' 
Educators ($n = {30}$) also detailed how the proctoring tools restrict a student's activities and access to unauthorized resources. 
Educator P41 (College of Arts \& Sciences) explained, ``The software takes control of a students computer so that they can't leave the exam, can't access the internet, and can't access other programs on the computer.'' 
Additionally, educators ($n = {19}$) described how proctoring tools record and flag anomalies during the exam taking session. 
For example, educator P114 (School of Public Health) added, ``Software can also record the user (visual and audio) while the user is taking the test and can flag any suspicious activity (user getting up from the computer, looking down at table, etc.) for the instructor to later review to determine whether cheating occurred.''
However, only a few ($n = {3}$) educators described the ability of the proctoring service to verify a student's identity.
P44 (College of Arts \& Sciences; Respondus) noted, ``There are various ways that they can monitor student identity while taking the exam (web cam).''

Some educators ($n = {32}$) reported that they did not know how they work. Educator P16 shared, ``I don't know anything about how they work.''
Others ($n = {10}$) had simply not used them or were not aware of there availability at the university.

\paragraph{Reasons for Not Using Online Proctoring Services}
Of the 79\% ($n={99}$) of educators responding to the survey that did not use online exam proctoring tools, 25\% ($n = {25}$ of ${99}$) considered doing so at some point (\ref{appendix:main-survey:N1}).
The College of Arts \& Sciences had highest percentage (30\%; $n = {12}$ of ${40}$) of educators who considered using exam proctoring, and likewise, S.T.E.M  (29\%; $n = {14}$ of ${49}$) was the highest percentage subject matter (see Tables \ref{tab:org-unit-table} and \ref{tab:subject-table}).
(\ref{appendix:main-survey:N2}). 
For instance, P77 (College of Arts \& Sciences) who selected \emph{Somewhat unlikely} to use online exam proctoring (\ref{appendix:main-survey:N5}) explained,
\begin{quote}
\small
Cheating during exams is a serious (and somewhat common) issue. Being assured that cheating was kept to a minimum would provide confidence that grades were well-earned.
\end{quote}

For many of the 73\% ($n = {72}$ of ${99}$) of educators who did not consider using online proctoring tools, their decision was likely informed by their negative perceptions of these tools \ref{appendix:main-survey:N3}). Many ($n = {24}$) reported that they considered online exam proctoring tools to be harmful to students. Educator P10 College of Professional Studies, \emph{Extremely unlikely}) stated, ``Proctoring tools monitor the students in ways that increase their anxiety and obliterate their ability to learn.'' Likewise, educator P77 (College of Arts \& Sciences, \emph{Somewhat unlikely}) shared,
\begin{quote}
\small
Privacy and home issues. Not all students were in a position to be engaged in being a student while remote learning. If a student was a primary care giver for a child or elderly person, how could I penalize them from looking away from their screen during an exam?
\end{quote}
P10 (College of Professional Studies, \emph{Extremely unlikely}) even described it as ``prison technology,'' and 
others ($n = {11}$) concluded that it impacts the trust between student and educator. For instance, educator P81 (College of Arts \& Sciences, \emph{Somewhat unlikely}) noted, ``It feels invasive and I feel it erodes trust between student and professor.'' Furthermore, educators ($n = {10}$) had concerns about the negative impact on student privacy, such as when educator P76 (College of Arts \& Sciences, \emph{Extremely unlikely}) shared, ``I found the on-line proctoring system to be a serious invasion of privacy of the student.''

Some ($n = {9}$) educators determined that online proctoring tools lacked the ability to actually stop cheating, like when P53 (College of Arts \& Sciences, \emph{Extremely unlikely}) said, ``\dots just locking down a computer is meaningless when students can easily access a second computer (or their phone).'' 
There were also a number ($n = {16}$) who were not aware of the availability of the tools at the university. Educator P16 (College of Arts \& Sciences, \emph{Neither likely nor unlikely}) noted, ``I have no knowledge of them or their availability.''

Many educators refactored their assessment formats to avoid the necessity of online exams (\ref{appendix:main-survey:N4}). 
A common tactic was to provide time limits enforced through existing learning management
software, such as Blackboard. 
For example, educator P96 (College of Arts \& Sciences, \emph{Extremely unlikely}) stated, ``I gave exams on Blackboard. They were timed, so that students would have limited time to look up answers.''
Some educators ($n = {14}$) reported changing their exams to open book and open note exams, such as educator P90 (College of Arts \& Sciences, \emph{Somewhat unlikely}) who explained, ``I ended up making everything open book so that I did not have to police anything.'' 
Others switched to take home assessments, like P14 (School of International Affairs, \emph{Somewhat unlikely}) who added, ``During Covid I made the quizzes take home and open book.''
Replacing the exams with other forms of assessment, such as projects and written papers, was another common theme ($n = {34}$). 
For example, educator P41 (College of Arts \& Sciences, \emph{Somewhat likely}) said, ``I decided to replace my exams (midterms and finals) with two-week take-home projects, with many scaffolded layers.''
Another example is educator P109 (School of Business, \emph{Somewhat likely}) who shared, ``I ditched the quizzes and tests, opting instead for graded homework and written papers.''
Still others ($n = {4}$) reduced the percentage of the overall course grade which would come from exams, like P93 (School of Engineering \& Applied Science, \emph{Neither likely nor unlikely}) who illustrated, ``The only way I actively managed it was 1. to put tremendous credit on the term project and 2. reduced credit for exams.''

Another common theme ($n = {9}$) among educators who chose not to use online exam proctoring tools was a belief that students would not cheat when asked to adhere to the university code of academic integrity. For instance, P76 (College of Arts \& Sciences, \emph{Extremely unlikely}) stated,
\begin{quote}
\small
I told students that I expected them to be adults, and to follow the university expectations of integrity and honesty. This was after I told them my opinions of the proctoring system to be an invasion of their privacy. They appreciated my opinion and cooperated with taking exams with honesty.
\end{quote}
Educator P100 Navy ROTC, \emph{Somewhat unlikely}) trusted students to follow the university honor code and noted, ``I reminded my students of the honor code and trusted them to follow it.''

Finally, it appears that most educators responding to the survey  ($n = {53}$ of ${99}$;~54\%) who do not currently use online exam proctoring tools in their classes reported they would be unlikely to use them if they were teaching remotely under similar circumstances to those of the 2020/21 academic year, while only 24\% ($n = {24}$ of ${99}$) said they were likely  (\ref{appendix:main-survey:N5}). This suggests that those who declined to use remote proctoring are unlikely to change their opinion of the technology.
When describing why they choose not to use online exam proctoring tools (\ref{appendix:main-survey:N3}), the participants who reported being \emph{Extremely unlikely} or \emph{Somewhat unlikely} to use the tools in the future (\ref{appendix:main-survey:N5}), more often described themes such as proctoring tools as potentially harmful to students (20 of 53 vs. 3 of 24), privacy concerns (10 of 53 vs. 0 of 24), tools do not stop cheating (9 of 53 vs 0 of 24), and trust students not to cheat (9 of 53 vs. 2 of 24).

\paragraph{Reasons for Using Online Proctoring Services}
Twenty-six educators responding to the survey reported using online proctoring services. We asked them what factors they considered when deciding to use these tools in an open response question~(\ref{appendix:main-survey:Q8}). These factors may include majority opinions; however, they may not capture minority opinions due to the small number ($n=26$) of educators who used these services. 

The most cited reason for using online proctoring tools is the convenience they offered ($n = {12}$). For many ($n = {7}$) this convenience was attributed to their familiarity with the proctoring tools either because they (or their colleague) have previously administered an online proctored exam using the tool ($n = {6}$).
For instance, P69 (College of Arts \& Sciences; Respondus) noted, ``Familiarity based on discussions with colleagues (who all used respondus or proctor exams themselves)'' and P25 (School of Nursing; Proctorio) added, ``Already using this product - the proctoring version for the online environment is called Examplify (w/ ExamSoft).''
Others ($n = {4}$) were influenced to use online proctoring due to their apparent popularity and recommendations from others, like P67 who simply said, ``It was popular.''
A few  ($n = {2}$) educators noted that these tools had been recommended to them by their institution, such as P26 (College of Arts \& Sciences; Respondus) who recalled, ``It was recommended by the school.''

Many educators ($n = {8}$) noted that their main motivation to use remote proctoring was out of a form of necessity. Specifically, most ($n = {6}$) indicated that they were required to use online proctoring by their department in an attempt to make assessment more uniform. 
For example, P33 (School of Medicine \& Health Sciences; RPNow) said,
\begin{quote}
\small
RPNow is the one used by my department. I don't believe that I have a choice of which online proctoring service to use. It is already set up in my courses for me.
\end{quote}
Or because they were administering a standardized test ($n = {2}$), like P25 (School of Nursing; Proctorio) who stated, ``Nursing students also take standardized exams via ATI - their proctoring service in an online environment is called Proctorio.''
Some educators ($n = {2}$) felt compelled to use exam proctoring due to the circumstances of the COVID-19 pandemic and remote learning, e.g., P36 (School of Nursing; Examsoft) who stated, ``Absence due to Covid exposure.''

We further queried educators who indicated that they used exam proctoring in open-responses regarding the benefits of using these tools~(\ref{appendix:main-survey:Q11}). The  most frequently ($n = {16}$) mentioned benefit was to to enforce exam rules or exam integrity as the primary benefit using online proctoring. Specifically, educators  indicate that the tools help prevent cheating ($n ={10}$), protect the integrity of remote exams ($n = {2}$), and ensure exam fairness, by limiting the benefit, or competitive edge students can gain from cheating ($n = {2}$).  
For example, P33 (School of Medicine \& Health Sciences; RPNow) describes how online exam proctoring tools deter cheating as a primary benefit:
\begin{quote}
\small
Even if not activated, students go through the [remote proctoring] system to take their exams, so they are under the impression that they are always being monitored. Ensures integrity of the exam without having to re-write questions to be open book.
\end{quote}

Educators also felt that a benefit of online proctoring was to enforce exam rules by verifying students' identities ($n ={1}$), and holding students accountable for any misconduct they may commit ($n = {2}$). As P123 (School of Medicine \& Health Sciences; RPNow) describes, ``[The tools] provide a permanent record of the student's behavior during an exam.''
Other educators ($n = {2}$) highlighted how remote proctoring tools allowed them to enforce other exam rules, like time limits ($n={1}$) and prohibit access to prohibited resources ($n ={1}$), e.g. P26 (College of Arts \& Sciences; Respondus) added, ``Ensuring that students don't use web resources to complete the test.''

Many ($n = {13}$) educators noted as a benefit how online proctoring tools offered additional flexibility. 
As ($n = {11}$) instructors explained, these tools made it easy for them to set up and grade their exams. 
While ($n = {5}$) respondents highlighted the tools’ general ease of use, e.g. P21 (School of Nursing; ProctorU) noted, ``Ease of use can be done online.'' 
Additionally, instructors found it allowed them to give proctored exams while complying with COVID safety precautions, such as when P118 (School of Nursing; Proctorio) stated, ``Convenience and safety during COVID.''

Finally, a handful of educators noted that online proctoring made it easier to manage their exams when compared to proctoring exams in person ($n = {2}$), that the proctored were easier to grade ($n = {1}$), and that they were convenient to use ($n = {2}$). One educator noted the flexibility of using multimedia content in their exams when online, and others $(n={2})$ noted that it also provides flexibility to students in selecting the time and environment for their exam. For instance, P35 (School of Medicine \& Health Sciences; Respondus) shared, ``It gave flexibility to the students to take the exam when convenient instead of at a set time.''

\paragraph{Drawbacks when Using Online Proctoring Tools}
We also asked educators who indicated they used online proctoring tools ($n=26$) the drawbacks of online proctoring~(\ref{appendix:main-survey:Q12}). These drawbacks may include majority opinions; however, they may not capture minority opinions due to the smaller number ($n=26$) of educators who used these services.

Most educators ($n = {20}$) identified at least one technology or usability issue they encountered while doing so. Chief amongst the drawbacks were technology glitches ($n={7}$), and system limitations ($n={17}$) that hindered students' or educators' ability to use the tools for their intended functions. They noted that these limitations impacted their ability to monitor students during exam time, to control students’ test environment and ensure academic integrity, and to conclusively identify cases where students had cheated.
A cited cause ($n={3}$) for these issues were limitations on students' computers to run the proctoring software or students' internet access.
When it came to connecting to the proctoring tools, instructors noted that some students either had unstable internet connections ($n={1}$), or had limited access to their exams due to being located in a different country ($n={1}$). In other cases, students' computers seemed to be the point of failure. In particular, instructors noted that some students' sometimes used older computers ($n={1}$). This meant that their machines would occasionally freeze when running the proctoring tools ($n={1}$), that they would not have webcams or microphones through which their exam session could be recorded, or that these input devices would fail to record ($n={1}$) while students took their exams.  As P44 (College of Arts \& Sciences; Respondus) explains, the software's lack of dependability posed a significant ``obstacle for students.'' 

Educators also noted drawbacks with respect to the privacy of their students.
Several ($n = {6}$) cited concerns for their students’ privacy. In particular, they noted wariness about third parties potentially collecting vendor data about their students ($n ={1}$), and that they found monitoring via video or audio recording to be privacy invasive ($n = {3}$). As P33 (School of Medicine \& Health Sciences; RPNow) explained they ``[Felt] uncomfortable seeing students' living situation and watching them while taking the exam.''  (We elaborate more on the privacy concerns in \autoref{sec:rq2}.)

Additionally, two educators also described drawbacks with respect to the interpersonal relationship with students that subjecting them to online proctoring can have, and that they ($n=3$) were personally discomforted with the use of video and audio recording to monitor exams and using that to actually identify cases of cheating.  As P98 (School of Medicine \& Health Sciences; Respondus) describes
\begin{quote}
\small
The [proctoring software] utilizing the camera is an invasion of privacy, often didn't work, and had the students so paranoid that they would email me to explain any movement they made. Plus, I realized that it would be difficult to ever prove anyone was actually cheating... I quit using the camera halfway through because of these problems.
\end{quote}

\figAgreement
\figLessLikelyToCheat
\figCatchCheating

\paragraph{Effectiveness of Exam Proctoring}
Educators who responded to the survey and indicated that they used online proctoring tools ($n=26$) were asked about the effectiveness of these tools at reducing cheating~(\ref{appendix:main-survey:Q19}). Responses were mixed when considering if exam proctoring tools reduced cheating. Eleven ($42\%$) educators who used online proctoring tools either {\em strongly agreed} or {\em somewhat agreed} that exam proctoring reduced cheating, while the same amount ($n=11$ of 26; 42\%) {\em strongly} or {\em somewhat disagreed}. Four ($15\%$) {\em neither agreed nor disagreed}. 

There was less confidence that the proctoring software would actually catch cheating~(\ref{appendix:main-survey:Q20}). Roughly a third ($n=9$ of 26; 34\%) of educators believed proctoring tools would catch cheating at least 50\% of the time. In contrast, nearly two-thirds ($n=16$ of 26; 61.5\%) believed it caught cheating up to 50\% of the time. Refer to \autoref{fig:lessLikelyToCheat} for full details. Moreover, only 38\% of educators stated that the tools reported cases of cheating~(\ref{appendix:main-survey:Q21}). See \autoref{fig:catchCheating} for more information.  

This suggests that educators found the software to be more successful in {\em deterring} cheating than in actually detecting or catching cheating. For instance, educator P25 (School of Nursing, Proctorio) {\em somewhat agreed} that exam proctoring reduced cheating but reported it catches cheating less than 25\% of the time~(\ref{appendix:main-survey:Q20}) and said, ``They don't prevent cheating - students can look up ways on the Internet for workarounds. But they do deter cheating.'' Likewise, educator P125 (School of Medicine \& Health Sciences; Respondus), who {\em somewhat disagreed} that exam proctoring reduced cheating, stated it catches cheating less than 25\% of the time~(\ref{appendix:main-survey:Q20}) and wrote, ``The video monitor and flagging is not great. Really doesn't prevent cheating, may just deter for a lot of people.''

Despite clearly different opinions on the effectiveness of exam proctoring tools at preventing and identifying cheating, most of the educators ($n=14$ of 26; 54\%) who used them reported that they either {\em strongly} or {\em somewhat} agreed that they were a good solution in responses to \ref{appendix:main-survey:Q25} (see \autoref{fig:agreeBarChart}).

In response to \ref{appendix:main-survey:Q12}, with respect to drawbacks, three educators noted the proctoring tools’ audio and video monitoring capabilities did not allow instructors to fully inspect students’ exam conditions during assessments.  Eleven respondents indicated the tools did not eliminate cheating, and when cheating was reported, three suggested that there were inconsistencies in the reports, causing them frustration, and at least one false positive and one false negative. Ultimately, this led two educators to suspect a subset of students had likely cheated, but to have been unable to conclusively identify which students had done so, despite their use of the proctoring tools.

\figLikelyAlluvium

\paragraph{Continued Use of Online Proctoring}
We were interested in further exploring the impact of the COVID-19 pandemic on the decision to use online exam proctoring tools.
We asked the educators who reported using online proctoring ($n=26$) if they would use online proctoring tools again if they were teaching remotely under similar circumstances to those of the 2020/21 academic year (\ref{appendix:main-survey:Q13}). 
Sixty-five percent ($n = {17}$ of ${26}$) said they were {\em likely} to use online proctoring tools again under those circumstances. 
While only 27\% ($n = {7}$ of ${26}$) said they were {\em unlikely}. 

We followed up by asking why they would or would not use online exam proctoring tools in such a situation \ref{appendix:main-survey:Q14}). 
Educators shared that it was either the next best option ($n = {3}$) or the only option ($n = {3}$) when in-person proctoring was not available. 
For instance, P47 (School of Nursing; Examsoft) shared, ``If we were unable to test in person, this would be our only option.''
For others ($n = {2}$) it was to maintain exam integrity.
As educator P25 (School of Nursing; Proctorio) highlighted, ``It's the only main way to control for academic integrity when not face-to-face.''
Educator P118 (School of Nursing; Proctorio) reported pandemic safety was the reason and said, ``If it is about being safe during a pandemic, I will use remote proctoring software every time.''

Next, we asked the same educators how likely would they be to use online proctoring tools again if conditions were similar to Fall 2021 when in-person learning resumed with masks and some hybrid options~(\ref{appendix:main-survey:Q15}). 
Only 35\% ($n = {9}$ of ${26}$) reported they were {\em likely} to continue to use online proctoring tools, while 46\% ($n = {12}$ of ${26}$;~54\%) reported they were {\em unlikely}.

Finally, we asked the same educators how likely would they be to use online proctoring tools again if they were teaching classes fully in-person without hybrid options~(\ref{appendix:main-survey:Q17}). 
We observed similar responses with  35\% ($n = {9}$ of ${26}$) reported they were {\em likely} to use the tools, and over half ($n = {14}$ of ${26}$;~54\%) reported they were {\em unlikely}. 
Detailed results can be found in \autoref{fig:likelyCircumstanceFullVirtualToHybridAlluvial} and \autoref{fig:figLikelyCircumstanceHybridToFullInPersonLearning}.

We followed up again by asking why they would use online exam proctoring in this situation.
Some educators simply shared that there was no longer any need to use online exam proctoring when in-person classes were taking place ($n = {2}$), or that they preferred traditional in-person exam proctoring ($n = {3}$).
For example, educator P124 (College of Arts \& Sciences; Respondus) noted, ``So if there is no concerns about pandemic, exams should definitely be in person.''
Others said they would still consider using online exam proctoring for missed exams ($n = {1}$) or other extenuating circumstances ($n = {1}$), and asynchronous quizzes($n = {1}$).
Some educators would continue to use online exam proctoring for reasons of flexibility ($n = {1}$), hybrid course offerings ($n = {1}$), or because they prefer online exam proctoring ($n = {1}$). 
Educator P56 who referred to flexibility said, ``I think it is a helpful option for times when holding exams virtually provides flexibility for students and faculty while still meeting course objectives and assessment standards.''

Our results suggest that many educators have used online exam proctoring as a temporary expedient to manage assessments during pandemic induced remote learning periods.
However, our results also suggest that a subset of educators will likely continue to use online proctoring as classes return to full in-person learning.

\subsection{RQ2: Privacy and Security Concerns }
\label{sec:rq2}

\figSoftwareConcern

\paragraph{Privacy Concerns} As with any application, there are possible privacy and security risks for users. 
We asked educators responding to the survey questions regarding privacy in relation to exam proctoring tools. One of the questions asked educators if they thought monitoring tools were an invasion of privacy (\ref{appendix:main-survey:Q22}). Of the educators that indicated using proctoring tools ($n = {26}$ of ${125}$), the majority either {\em strongly disagreed} ($n = {6}$ of ${26}$) or {\em somewhat disagreed} ($n = {10}$ of ${26}$) with the concept of a proctoring service being an invasion of privacy. When the educators were separately asked if they were specifically concerned about privacy risks to students (\ref{appendix:main-survey:Q24}), the majority either {\em somewhat agreed} ($n = {8}$ of ${26}$) or {\em  strongly agreed} ($n = {4}$ of ${26}$). See \autoref{fig:agreeBarChart} for full results.

When asked to elaborate about the specific factors that informed their views about the privacy of online exam proctoring tools in an open response question (\ref{appendix:main-survey:Q26}), one common theme ($n = {9}$) among some of the educators who chose to use online exam proctoring tools was a belief that the privacy risks was acceptable. For instance, P28 (School of Medicine \& Health Sciences; Respondus) stated, 
\begin{quote}
\small
Most digital tools have some level of privacy issues. Any program that can access an internal mic or video seems to be a privacy risk. There are other learning tools such as Voice thread that I believe have a privacy risk but see the benefit outweighing the risk and damage to the student. 
\end{quote}
In contrast, many educators ($n = {9}$) expressed  discomfort with online exam proctoring tools. For example, educator P36 (School of Nursing; Examsoft) stated, 
\begin{quote}
\small
Any program that requires you to download a file, disables your system functionality, and automatically searches your computer for files to upload is a total invasion of privacy. Additionally, the proctoring service records the student at their most vulnerable- in their home environment where they sometimes forget they are being recorded. This leaves the potential for private matters being recorded and permanently on a server somewhere... if there is a data breach of this program, these videos could be out there for anyone to see.
\end{quote}

Some participants ($n = {3}$) called out the webcam as being privacy invasive, like P125 (School of Medicine \& Health Sciences; Respondus) who illustrated, ``Video monitoring is more intrusive.'' 
Educator P58 (College of Arts \& Sciences; Respondus) stated that artificial intelligence used to monitor student behavior in private was invasive when they said ``AI required to monitor and flag student behavior use of recordings in private setting.''
And educator P49 (College of Arts \& Sciences; Respondus) added, ``Glad to see the tradeoff question: Yes, it is somewhat invasive but that is offset by exam integrity.''

When asked whether they thought the remote exam proctoring tool they used offered a reasonable tradeoff between student privacy and exam integrity (\ref{appendix:main-survey:Q23}), respondents appeared to be hesitant to endorse the tools. Here, only a plurality of participants ($n = {12}$) indicated they {\em agreed} with the statement, where only ($n = {6}$) {\em strongly agreed} with the statement, and ($n = {6}$) {\em  somewhat agreed}. 
See \autoref{fig:agreeBarChart} for full results.

\paragraph{Software Security Concerns}

Exam proctoring services often require students to install specialized software to enable proctoring. 
The required proctoring software is often in the form of a browser extension that is added to students' existing web browsers or standalone software that must be installed on students' personal computers.
As is the case with any custom software, there is risk of security vulnerabilities.

We asked educators responding to the survey who had used remote proctoring tools how concerned they were about students installing software created by exam proctoring companies on their personal computers (\ref{appendix:main-survey:Q27}).
Over half of educators ($n = {14}$ of ${26}$;~54\%) were {\em not at all concerned}, while 31\% ($n = {8}$ of ${26}$) where {\em slightly} or {\em somewhat concerned}. Only 15\% ($n = {4}$ of ${26}$) were {\em moderately} or {\em extremely concerned}. Refer to \autoref{fig:softwareConcern} for the full results.

We then asked respondents to explain which factors led to their concern, or lack of concern, regarding students installing exam proctoring software (\ref{appendix:main-survey:Q27}). 
A number of educators ($n = {8}$) voiced concerns about the software.
Concerns such as reliability issues ($n = {3}$), potential invasion of student privacy ($n = {2}$), security flaws ($n = {1}$), negative impacts on computer functionality ($n = {1}$). 
Educator P36 (School of Nursing; Examsoft) share concerns about privacy, ``Again, any program that requires you to download a file, disables your system functionality, and automatically searches your computer for files to upload is a total invasion of privacy.''
Whereas, educator P124 (College of Arts \& Sciences; Respondus) considered tradeoffs between privacy and necessity when the educator said, 
\begin{quote}
\small
No one likes to install software on their computer that could potentially be invasive. It's necessary in this instance but I can see why someone would be reluctant to do so.
\end{quote}
Still a number of educators ($n = {6}$) did not have concerns, such as educator P28 (School of Medicine \& Health Sciences; Respondus) who noted, ``We install so much on our devices so I don’t see this as a higher risk than other applications.''
We also found statements describing a transfer of trust from the institution, which licensed the software and made it available to educators, to the exam proctoring software itself. 
This implied trust leads educators to assume that the software has been through a vetting process. 
For instance, educator P125 (School of Medicine \& Health Sciences; Respondus) responded, ``If recommended by University then assume it is safe.'' 
And educator P25 (School of Nursing; Proctorio) added, ``Just don't know enough to answer this question; defer to our [Online Learning and Instructional Technology] team who vet the software.''

\subsection{RQ3: Proctoring Methods}
\label{sec:rq3}

\figMonitoringMethodsEnabled

\paragraph{Enabling Monitoring Methods} 
Online exam proctoring services provide numerous types of student monitoring methods. 
These monitoring techniques range from lockdown browsers that prevent navigation to other sites during exam time, to more invasive monitoring that may include webcams, screen sharing, the use of a live (human) proctor, and even automated monitoring techniques such as eye tracking and network traffic analysis. 
Educators must select the monitoring methods they deem appropriate for proctoring students while they complete assessments. 

We asked educators to report all of the monitoring methods they enabled in their proctored exams (\ref{appendix:main-survey:Q29}).
Most educators ($n = {21}$ of ${26}$;~81\%) reported enabling the lockdown browser, which many educators find to be the least invasive proctoring technique. 
Fifty percent of the educators ($n = {13}$ of ${26}$) enabled webcam recording, and many educators ($n = {11}$ of ${26}$;~42\%) enabled microphone recording and face detection. 
Still others ($n = {8}$ of ${26}$;~31\%) enabled the arguably more invasive monitoring methods of screen recording and eye movement tracking.
Refer to \autoref{fig:monitoringTypesWithComfortMonitoringStudentsBar} for the full results.

\paragraph{Monitoring Method Effectiveness}
A majority ($n = {16}$ of ${26}$;~62\%) of educators reported that they would enable the same monitoring methods again to administer another online proctored exam (\ref{appendix:main-survey:Q30}). 
While 38\% ($n = {10}$ of ${26}$;) reported they would not use ($n = {3}$ of ${26}$;~11\%), or were unsure if they would use ($n = {7}$ of ${26}$;~27\%), the same monitoring methods. 

When asked what monitoring methods in their online proctored assessments they would change and why (\ref{appendix:main-survey:Q31}). 
Educator P19 (College of Arts \& Sciences; Respondus), who wanted to remove the webcam monitoring said, ``Students did not feel comfortable or said they did not have a camera so we could not go that route.'' 
Furthermore, educator P124 (College of Arts \& Sciences; Respondus) who wanted to remove the face detection monitoring shared, ``Facial detection not as necessary I don't turn on the option for Respondus to fire off warnings when students face disappear from view, but I do watch the recordings later to determine if there is any egregious violations.''

Additionally, educators reported technology issues with monitoring that relies on the webcam or microphone. For instance, educator P47 (School of Nursing; Examsoft) stated, 
\begin{quote}
\small
There is no way of guaranteeing that the student's webcam and microphone are working during the test. It is not until after that we can determine if they were working and by then, it's too late.
\end{quote}
Bandwidth and lack of staff to view the videos could also be an issue as educator P103 (School of Business; Respondus) added, ``Most students had excuses not to have cameras, the low bandwidth was a problem with Respondus, and we don't have enough staff for watching/proctoring.''

\paragraph{Comfort With Monitoring Methods} 
We asked educators how comfortable they feel using each monitoring type during an online proctored exam (\ref{appendix:main-survey:Q33}).
Overall educators were comfortable with all 12 monitoring types presented to them. 
The lockdown browser monitoring method had the largest number ($n = {22}$ of ${26}$;~85\%) of educators who reported being {\em comfortable}, many $n = {20}$ of ${26}$;~77\%) of them {\em extremely comfortable}. 
This is followed by internet activity monitoring at 65\% ($n = {17}$ of ${26}$) of educators {\em comfortable} and keyboard restrictions at 54\% ($n = {14}$ of ${26}$). 
Please refer to \autoref{fig:monitoringTypesWithComfortMonitoringStudentsBar} for the full results.
These results are notably inline with the results found by Balash \etal when they asked students to select their comfort level with online exam proctoring monitoring methods\cite{balash-21-exam}.

\figConcernStudentsSharingInformation

\paragraph{Information Sharing Concern} 

We asked educators to report their level of concern on a 5-point Likert concern scale for students sharing these various types of information with exam proctoring companies (\ref{appendix:main-survey:Q34}).
Educators were generally {\em unconcerned} with most types of student information sharing, except for identifiable information such as social security number, date of birth, street address and location data. 
The student information that garnered the largest number ($n = {22}$ of ${26}$;~85\%) of {\em concerned} educators was social security number.  
Please refer to \autoref{fig:concernStudentsSharingInformationBar} for the full results.

\paragraph{Exploring Factors Influencing Proctor Usage} 

We performed exploratory analysis to determine potential factors that could lead to increased/decreased usage of exam proctoring in the future. As we had no priors and a small sample, we applied feature reduction analysis using multiple logistic regressions, considering all possible factors and reducing individual until the model either did not converge or the variance was no longer improving. 
As this exploratory, we refrain from presenting odds ratios and p-values in text and instead focus on factors that could be explored more in future research.
The factors we considered included educator comfort with the 12 exam monitoring types (\ref{appendix:main-survey:Q33}), concern for 11 student information sharing types (\ref{appendix:main-survey:Q34}), agreement that online exam proctoring tools makes it less likely that students will cheat (\ref{appendix:main-survey:Q19}), the percentage of actual cheating found (\ref{appendix:main-survey:Q20}), and if the tool reported potential cheating (\ref{appendix:main-survey:Q21}).
As outcomes we considered the likelihood of using online exam proctoring tools for assessments assuming similar circumstances to the 2020/2021 academic year (\ref{appendix:main-survey:Q13}), Fall 2021 (\ref{appendix:main-survey:Q15}), and a full return to in person learning (\ref{appendix:main-survey:Q17}). The outcomes were binned into two levels, educators who were \emph{Extremely Likely} or \emph{Somewhat Likely} to use proctoring and those who were not. 
The full models are found in  \autoref{appendix:additional}. 

The leading factors that survived reduction were visible proctors, sharing student info, and does proctoring actually prevent cheating. 
For instance, we found (\autoref{table:likely-fallspring-regression}) a correlation with participants who were \emph{Extremely Comfortable} or \emph{Somewhat Comfortable} with a live proctor not visible to students and internet monitoring of students (\ref{appendix:main-survey:Q33}). Those participants were more likely to use online exam proctoring tools given similar circumstances to those of the 2020/2021 academic year (\ref{appendix:main-survey:Q13}). And we find a similar correlation for those who \emph{Strongly Agree} or \emph{Somewhat Agree} that the use of online exam proctoring tools makes it less likely that students will cheat (\ref{appendix:main-survey:Q19}).
Future work could design experiments around these factors to quantify the effects.

\section{Discussion and Conclusions}\label{sec:discussion}

We surveyed $(n=125)$ educators at a large academic institution about their perceptions and use of online proctoring during the 2020/21 academic year. Of those who responded to the survey, most ($n=99$; 79\%) did not use online proctoring, with some arguing that it was invasive or unnecessary. 

Of those that did use online proctoring tools ($n=26$; 21\%), many felt that they did not have a choice  either because of the   necessity to maintain academic integrity or because they were required to do so by their department. Educators who used online proctoring were also not uniform in their view that it actually helps to deter and detect cheating, despite noting that it was a good solution under the circumstances. Furthermore, there was general comfort among educators who used remote proctoring with the monitoring of students (e.g., via video, screen share, or microphone); these educators were more concerned with sharing student data (e.g., name, student identification, etc.) with online proctoring companies. %

Moving forward, even in a situation where there is full, in-person learning, many educators that use online proctoring indicated they would continue to do so, suggesting that more work is needed to address the potential privacy and security risks for students and educators when using these tools. %

\paragraph{Privacy Tradeoffs}
There were marked differences between the educators who used online exam proctoring and those who chose not to use the tools.
Educators who did not use online proctoring tools can generally be classified into one of two categories:
The first consists of educators who preferred to redesign their assessments so they could be more easily completed remotely without concern for academic integrity violations, such as open book/note/internet exams and writing or project-based assignments. 
The second consist of those who considered the tradeoffs between student privacy and the utility of the tools and decided that the potential privacy and security risks to student test takers outweighed the utility of the tools. 
Overall, the instructors who did not use remote proctoring had the most thematically negative responses to these tools and often highlighted the privacy risks and potential harms to students. 

Likewise, the educators in our study that did use online proctoring tools can generally be classified into one of two categories:
The first category are those who were forced to use online proctoring services either by their department, organizational unit, or as a standardized testing requirement. Their opinions of the tools generally better matched those who chose not to use them, but generally thought of them as less harmful and privacy invasive than those who did not use online proctoring at all.
The second category is educators who considered the tradeoffs between student privacy and the utility of the tools and decided that the need for academic integrity outweighed the potential privacy and security risks to student test takers.

\paragraph{Educator Training and Guidance for Online Proctoring}
Many educators in the study expressed a desire for training to better understand the available proctoring tools and their impacts. 
Overall, they demonstrated general knowledge about how exam proctoring tools function with respect to monitoring students  and how they restrict the use of unauthorized resources.  
However, educators seemed unaware of the methods used to validate students' identities and what happens to students' information after its collection for this purpose.

This presents an opportunity for improved training and guidance at the institutional level that provides the pros and cons about online proctoring and associated privacy/security risks.
Such training and guidance could also include  technical details on how exam proctoring tools can help educators maintain principles of least monitoring by using the smallest number of monitoring types necessary, given the constraints of the class. Moreover, institutional involvement in such training could set clear recommendations to set expectations for both educators and students.

\paragraph{Limitations on Enforcing Academic Integrity} 

Qualitative responses suggest that educators have broad skepticism of the limitations about what remote proctoring tools can actually do to ensure exam integrity. 
Many often highlighted the difference between the ability to deter potential cheating versus the inability to detect motivated cheating. 
For instance, online exam proctoring tools may prevent panic cheating, but they will not necessarily stop or detect more planned or sophisticated cheating techniques, such as secondary devices, virtual machines, or other workarounds. %
Furthermore, even educators who use the tools are fairly split on whether online proctoring actually deters and detects cheating.

\paragraph{Transfer of Trust}
In the qualitative responses from educators, we find evidence of a transfer of trust between the institutions who licence and provide the online exam proctoring software and the software itself. 
A similar finding was reported by Balash et al.~\cite{balash-21-exam} when surveying students on their opinions of online proctoring, noting that students trusted the institution and since the institution licensed online proctoring, they implicitly trusted online proctoring tools.
We found that the educators who used online proctoring expressed the same sentiment. They believed that their institution would not provide the software to proctor exams if it was not safe for students to install on their computers. 

Institutional support for third-party proctoring software, which conveys credibility, makes the exam proctoring software appear safer and less potentially problematic because educators assume that institutions have properly vetted the software and the methods used by the proctoring services. 

It is unclear that such trust is warranted.  All software has inherent risks of security vulnerabilities, and recent major security and privacy incidents have shown that online exam proctoring software has been subject to both major data breaches and to security vulnerabilities that allowed remote activation. 
Given the capabilities of exam proctoring software to monitor users and disable system functionality, extra precautions should be taken to reduce security risks to students who are required to install the software.

\paragraph{Implications to Security and Privacy}
Remote proctoring systems are naturally invasive.  When operating correctly, 
they monitor and restrict how students can interact with their own computers.  The consequences of security vulnerabilities and breaches (cf.~\cite{bleepingProctorUBreach2020,Proctorio2021ndl,burgess2022watching,chroniclesProctorio2022}) are significant, especially given the private information collected about students---for example, their physical locations, photos and videos of their environment, and information about their computing devices.

Understanding how and why users choose (or are forced to use) software that could harm their privacy and security is a critical research need.  This paper examines the perceptions of the decision-makers who choose whether or not to require remote proctoring---a form of monitoring software that has seen explosive growth.  As argued above, providing additional guidance and training to educators, and heightening their awareness of the privacy and security risks that these systems impose on their students, is paramount.  More generally, as with other technologies that aim at restricting and monitoring user functionality (e.g., remote IT management software), an argument can be made that limiting users' ability to control their own devices is antithetical to security and privacy.  Given the proliferation of remote proctoring, there is an urgent need to better understand not only the potential for abuse and misuse of these systems, but also the perceptions of the educators who have the power to decide if and how they are used.

\subsubsection*{Data and Source Availability}
All data, scripts, and qualitative codebooks are available at the following repository: \url{https://github.com/gwusec/2023-USENIX-Educator-Perspectives-of-Exam-Proctoring}
\subsubsection*{Acknowledgements}
We thank the anonymous shepherd and reviewers for improving this manuscript and preparing it for publication. This material is based upon work supported by the National Science Foundation under Grant Nos. 1845300, 2138654, and 2138078. 

\begin{footnotesize}
\bibliographystyle{plain}
\bibliography{bib/main.bib}
\end{footnotesize}

\appendix
\section{Survey Instrument}\label{sec:appendix}
\label{appendix:main-survey}
\setlength{\columnsep}{0pt}
\setlength{\multicolsep}{0pt}

\begin{questions}[label=\textbf{Q\arabic*}]

    \item Are you either a full-time faculty or part-time/adjunct faculty member at GW?
    \label{appendix:main-survey:Q1}
    \begin{multicols}{2}
        \begin{answers}
            \item Yes
            \item No
        \end{answers}
    \end{multicols}

    \item What is the primary organization unit or school you are associated with at GW?
    \label{appendix:main-survey:Q2}
    \begin{answers}
        \item {[\emph{List of 14 schools or organizational units}]}
        \item Prefer Not To Disclose
        \item Other (Please Specify) \hrulefill
    \end{answers}

    \item What subject area(s) (e.g., Math, English) did you teach during the 2020/2021 academic year? Enter ``N/A'' if you prefer not to disclose this information.
    \label{appendix:main-survey:Q3}

    \item These are the major online exam proctoring tools available at GW: For each tool note whether you were aware of its availability during the 2020/2021 Academic Year.
    \label{appendix:main-survey:Q4}
    \begin{tabularx}{\linewidth}{@{}l@{}C{1}@{}C{1}@{}C{1}@{}}
        \toprule
        \multicolumn{4}{r}{Aware~~Unaware~~I'm Unsure} \\
        \midrule
        Examity & $\bigcirc$ & $\bigcirc$ & $\bigcirc$ \\
        Examsoft & $\bigcirc$ & $\bigcirc$ & $\bigcirc$ \\
        Honorlock & $\bigcirc$ & $\bigcirc$ & $\bigcirc$ \\
        Proctor Track (Verificent) & $\bigcirc$ & $\bigcirc$ & $\bigcirc$ \\
        ProctorU & $\bigcirc$ & $\bigcirc$ & $\bigcirc$ \\
        Question mark & $\bigcirc$ & $\bigcirc$ & $\bigcirc$ \\
        Respondus & $\bigcirc$ & $\bigcirc$ & $\bigcirc$ \\
        Software Secure PSI (Remote Proctor Now) & $\bigcirc$ & $\bigcirc$ & $\bigcirc$ \\
        \bottomrule
    \end{tabularx}

    \item To the best of your understanding, how do online exam proctoring tools work?
    \label{appendix:main-survey:Q5}
    
    \item Did you use any online exam proctoring tools to assist in administering any exams or assignments during the Fall 2020/Spring 2021 academic year?
    \label{appendix:main-survey:Q6}
    \begin{multicols}{3}
        \begin{answers}
            \item Yes
            \item No
            \item I'm Unsure
        \end{answers}
    \end{multicols}
\begin{center}\vspace{-.15in}\textit{{\color{red} [\ref{appendix:main-survey:N1}-\ref{appendix:main-survey:N5} only if \ref{appendix:main-survey:Q6} is No or I'm Unsure]}}\end{center}\vspace{-1em}
\begin{questions}[label=\textbf{N\arabic*}]

    \item Did you consider using any online exam proc. tools?
    \label{appendix:main-survey:N1}
    \begin{multicols}{3}
        \begin{answers}
            \item Yes
            \item No
            \item I'm Unsure
        \end{answers}
    \end{multicols}
    
    \item Why did you consider using online exam proc. tools?
    \label{appendix:main-survey:N2}
    
    \item Why did you ultimately choose not to use online exam proctoring tools?
    \label{appendix:main-survey:N3}

    \item For course assessments, like exams, quizzes, etc., how did you manage student academic integrity for online classes during the 2020/2021 academic year? Please explain your decision making process \\
    \label{appendix:main-survey:N4}
    
    \item If you were teaching remotely under similar circumstances to those of the 2020/2021 academic year, and you decided to administer an exam online, how likely would you be to use online exam proctoring tools? 
    \label{appendix:main-survey:N5}
        \begin{answers}
            \item {[\emph{Five-point Likert likelihood scale}]}
        \end{answers}
\end{questions}
\vspace{-.2in}
\begin{center}\textit{{\color{red} [\ref{appendix:main-survey:Q7}-\ref{appendix:main-survey:Q34} are shown only if answer to \ref{appendix:main-survey:Q6} is Yes]}}\end{center}\vspace{-1em}
    \item Which online proctoring service did you use most recently?
    \label{appendix:main-survey:Q7}
        \begin{answers}
            \item {[\emph{List of online proctoring services from \ref{appendix:main-survey:Q4}}]}
        \end{answers}

    \item Which factors did you consider when making your decision to use online proctoring services?
    \label{appendix:main-survey:Q8}

    \item What kind of assessment did you administer most recently using an online exam proctoring service?
    \label{appendix:main-survey:Q9}
    \begin{answers}
        \item Course quiz
        \item Course exam (e.g. test, midterm exam, final exam)
        \item I have not administered an exam with online proc. %
        \item Other (Please Specify) \hrulefill
    \end{answers}
    
    \item How many online proctored assessments have you administered?
    \label{appendix:main-survey:Q10}
    \begin{multicols}{6}
        \begin{answers}
            \item 0
            \item 1
            \item 2
            \item 3
            \item 4
            \item 5+
        \end{answers}
    \end{multicols}

    \item In your experience, what were the main benefits of using online exam proctoring?
    \label{appendix:main-survey:Q11}

    \item In your experience, what are the main drawbacks of using online exam proctoring?
    \label{appendix:main-survey:Q12}

    \item If you were teaching remotely under similar circumstances to those of the 2020/2021 academic year, and you decided to administer an exam online, how likely would you be to use online exam proctoring tools?
    \label{appendix:main-survey:Q13}
        \begin{answers}
            \item {[\emph{Five-point Likert likelihood scale}]}
        \end{answers}
        
    \item Why would or wouldn't you consider using online exam proctoring?
    \label{appendix:main-survey:Q14}
    \item Based on your prior experience with online exam proctoring, in the current teaching environment at GW during Fall 2021, how likely are you to use online exam proctoring tools for these online assessments?
    \label{appendix:main-survey:Q15}
        \begin{answers}
            \item {[\emph{Five-point Likert likelihood scale}]}
        \end{answers}

    \item Why are, or aren't you considering administering assessment online during Fall 2021?
    \label{appendix:main-survey:Q16}

    \item Based on your prior experience with online exam proctoring, \textbf{assuming the end of the pandemic and a full return to in person learning}, how likely would you be to use online exam proctoring tools for these assessments?
    \label{appendix:main-survey:Q17}
        \begin{answers}
            \item {[\emph{Five-point Likert likelihood scale}]}
        \end{answers}
    
    \item Why would, or wouldn't you consider using online exam proctoring if in person learning resumed?
    \label{appendix:main-survey:Q18}

\paragraph{Understanding of Functions}
In this part of the survey you will be asked about the functionality of online exam proctoring services.

    \item The use of online exam proctoring tools makes it less likely that my students will cheat on an exam.
    \label{appendix:main-survey:Q19}
        \begin{answers}
            \item {[\emph{Five-point Likert agreement scale}]}
        \end{answers}

    \item I believe that online proctoring will catch cheating this percent of the time when students are actually cheating:
    \label{appendix:main-survey:Q20}
    \begin{multicols}{2}
        \begin{answers}
            \item 0-25\%
            \item 26-50\%
            \item 51-75\%
            \item 76-100\%
            \item Prefer not to answer
        \end{answers}
    \end{multicols}

    \item Did [Exam Proctoring Tool] report any potential cheating during an exam?
    \label{appendix:main-survey:Q21}
    \begin{multicols}{2}
        \begin{answers}
            \item No
            \item Yes
            \item I'm Unsure
            \item Prefer not to answer
        \end{answers}
    \end{multicols}
    
You indicated above that you used [Exam Proctoring Tool] recently to administer an online exam. Please indicate your level of agreement with the following statements based on your experience with [Exam Proctoring Tool].
    
    \item I think [Exam Proctoring Tool] is privacy invasive.
    \label{appendix:main-survey:Q22}
        \begin{answers}
            \item {[\emph{Five-point Likert agreement scale}]}
        \end{answers}

    \item I think [Exam Proctoring Tool] offers a reasonable tradeoff between student privacy and the integrity of the exam.
    \label{appendix:main-survey:Q23}
        \begin{answers}
            \item {[\emph{Five-point Likert agreement scale}]}
        \end{answers}

    \item I am concerned about the risks to student privacy by using [Exam Proctoring Tool].
    \label{appendix:main-survey:Q24}
        \begin{answers}
            \item {[\emph{Five-point Likert agreement scale}]}
        \end{answers}
    
    \item I think [Exam Proctoring Tool] is a good solution for monitoring remote examinations.
    \label{appendix:main-survey:Q25}
        \begin{answers}
            \item {[\emph{Five-point Likert agreement scale}]}
        \end{answers}

    \item Which specific factors inform your views about the privacy of [Exam Proctoring Tool]?
    \label{appendix:main-survey:Q26}
    
    \item How concerned are you about students installing software created by [Exam Proctoring Tool] on their personal computers?
    \label{appendix:main-survey:Q27}
        \begin{answers}
            \item {[\emph{Five-point Likert concerned scale}]}
        \end{answers}

    \item Which factors led to your concern, or lack of concern, about students installing [Exam Proctoring Tool] software?
    \label{appendix:main-survey:Q28}

\paragraph{Methods}
In this part of the survey you will be asked about the methods employed by online exam proctoring services.
 
Previously, you indicated that you administered an online exam via [Exam Proctoring Tool] offered at GW. Please refer to that experience in answering the following questions.

    \item Select all monitoring methods you enabled in your proctored exams. [Select all that apply]
    \label{appendix:main-survey:Q29}
    
    [\emph{Refer to \autoref{fig:monitoringTypesWithComfortMonitoringStudentsBar} for the list of monitoring methods}]

    \item If you were to administer another online proctored exam would you enable the same monitoring options again?
    \label{appendix:main-survey:Q30}
    \begin{multicols}{3}
        \begin{answers}
            \item No
            \item Yes
            \item I'm Unsure
        \end{answers}
    \end{multicols}
    
    \item You indicated that you used the following monitoring methods in your online proctored assessments: [Monitoring Methods]
Please explain why you felt these monitoring options worked well, or why they would work well proctoring a future exam.
    \label{appendix:main-survey:Q31}

    \item What monitoring methods in your online proctored assessments would change and why?
    \label{appendix:main-survey:Q32}

    \item For each exam monitoring type please select how comfortable you would feel about using it to monitor students during online proctored exams in your course.
    
    [\emph{Five-point Likert comfort scale}]
    
    [\emph{List of monitoring types from \ref{appendix:main-survey:Q29}}]
    \label{appendix:main-survey:Q33}

\paragraph{Privacy Concerns} 
In this part of the survey you will be asked about the benefits and potential risks you associate with online exam proctoring.

    \item Many online proctoring tools require students to share information with the exam proctoring companies. For each type of information below indicate whether you would be concerned by students sharing this information with \textbf{exam proctoring companies.} 
    \label{appendix:main-survey:Q34}

    [\emph{Participants selected from a five-point Likert concern scale from "Not at all Concerned" to "Extremely Concerned" for each information type}]

    [\emph{Refer to \autoref{fig:concernStudentsSharingInformationBar} for the list of information types}]

\paragraph{End of Survey}
We thank you for your time spent taking this survey. Your response has been recorded.
    
\end{questions}

\onecolumn
\section{Additional Figures and Tables}\label{appendix:additional}
\likelySpringFallRegression
\likelyFallRegression
\likelyFullReturnRegression

\figConcernStudentsSharingInformationExpanded


\twocolumn

\section{Qualitative Codebook}\label{appendix:codebook}
\begin{itemize}
\item\textbf{monitor-students (65)}

\emph{webcam (31), track-behavior (18), record-session (12), track-eye-movement (8), track-environment (4), prevent-cheating (3), verify-identity (3), keystrokes (2), flag-anomalies (2), log-activities (2), track-location (1), tracks-IP (1), microphone (1)}

\item\textbf{do-not-know (32)}

\emph{unaware (6)}

\item\textbf{restrict-activities (30)}

\emph{lockdown-browser (12), prevent-cheating (1)}

\item\textbf{harmful-to-students (24)}

\emph{invasive (7), intrusive (3), increased-stress (3), false-positives (2), increased-anxiety (1), decrease-learning (1), confusion (1), fairness-concerns (1), fear (1)}

\item\textbf{timed-tests (21)}

\emph{one-q-at-a-time (2)}

\item\textbf{flag-anomalies (19)}

\item\textbf{unaware (19)}

\emph{did-not-pay-attention (1)}

\item\textbf{written-papers (17)}

\item\textbf{open-book-exam (14)}

\item\textbf{NA (12)}

\item\textbf{in-person-exams (11)}

\item\textbf{trusts-students (11)}

\item\textbf{privacy-concerns (11)}

\item\textbf{open-book (11)}

\item\textbf{does-not-eliminate-cheating (11)}

\emph{students-cheated (2), prohibited-materials (2), second-computer (1), workarounds (1), smart-phones (1), physical-notes (1)}

\item\textbf{not-consider (11)}

\emph{prefer-in-person (3), cheating-too-easy (3), no-need (2), student-privacy (1), more-equity-in-person (1)}

\item\textbf{stop-cheating (10)}

\emph{use-web-resources (5), device-use (1)}

\item\textbf{cheating (10)}

\emph{caught-students (3), reduce (2), reduce-cheating (1)}

\item\textbf{do-not-use (10)}

\item\textbf{possible-exam-method (9)}

\emph{convenient (1), only-option (1), used-previously (1), preferred-online (1), midterm (1), sub-in-person-proctor (1)}

\item\textbf{no-exams (9)}

\item\textbf{acceptable (9)}

\emph{benefit-outweighs-risk (3), if-lockdown-only (2), awkward-but-necessary (1), request-video-deletion (1), monitored-in-person (1), some-privacy-issues (1), used-as-student (1)}

\item\textbf{honor-code (9)}

\item\textbf{discomfort (9)}

\emph{remote-data-storage (3), recorded-when-vulnerable (2), not-worth-risks (2), searches-files (1), ineffective-tool (1), assume-student-guilt (1), controls-computer (1)}

\item\textbf{does-not-prevent-cheating (9)}

\item\textbf{unchanged (9)}

\emph{deters-cheating (4), resource-restrictions (4), webcam (3), lockdown-browser (3), methods-provide-evidence-of-student-activity (1), zoom-proctoring (1), eye-movement-tracking (1)}

\item\textbf{concern (8)}

\emph{reliability-issues (3), invasion-of-privacy (2), connectivity-issues (1), forcing-students-to-use (1), requires-download (1), ways-to-get-around-protections (1), searches-computer-for-files (1), disables-system-functionality (1), security-software (1)}

\item\textbf{unsure-how-to-use (8)}

\item\textbf{application-based-questions (8)}

\item\textbf{projects (7)}

\item\textbf{blackboard (7)}

\emph{open-book (2), assignments (1), proctored-exam (1), safe-assign (1)}

\item\textbf{technology-glitches (7)}

\emph{camera (2), program-working (1), causes-stress-for-students (1), VPNs (1), old-device (1), computer-freezes (1), microphone (1), poor-internet-connection (1), causes-stress-for-instructor (1)}

\item\textbf{only-option-available (7)}

\emph{on-blackboard (3), program-decided-to-standardize (1), proctorio (1)}

\item\textbf{unsure (7)}

\emph{what-happens-to-collected-data (1), defer-to-admin (1), of-installation-process (1), of-any-malware (1)}

\item\textbf{would-consider (6)}

\emph{prefer-online-proctoring (1), for-quizzes (1), in-hybrid-courses (1), offers-flexibility (1), ensure-test-validity (1)}

\item\textbf{written-based-exams (6)}

\item\textbf{student-privacy (6)}

\emph{third-party-access-to-data (1), camera (1), monitoring (1)}

\item\textbf{already-using-online-proctoring (6)}

\emph{department-using (2), RPNow (1), examplify (1), proctorio (1), nursing-students-standardized-exams (1), Respondus-Monitor (1)}

\item\textbf{required-to-use (6)}

\emph{department (3)}

\item\textbf{lack-of-concern (6)}

\emph{it-worked-for-most-students (1), only-used-when-accessing-the-exam (1), have-read-privacy-policy (1), not-higher-risk-than-other-applications (1), recommended-by-university (1), widespread-use (1)}

\item\textbf{continue-remote-proctoring (5)}

\emph{final-exam-date-late (1)}

\item\textbf{trust-in-students (5)}

\item\textbf{safe-assign (5)}

\item\textbf{project-based-exams (5)}

\item\textbf{does-not-trust-tools (5)}

\emph{langauge-exam (1)}

\item\textbf{ease-of-use (5)}

\emph{dont-have-to-write-open-book-exam (2), delivery (1), setup (1), flexible-timing (1)}

\item\textbf{no-privacy-issue (4)}

\item\textbf{did-not-have-time (4)}

\item\textbf{may-consider (4)}

\emph{depends-on-students (1), asynch-quizzes (1), extenuating-circumstances (1), for-missed-exams (1)}

\item\textbf{tests-worth-little (4)}

\item\textbf{sign-integrity-code (4)}

\item\textbf{none (4)}

\item\textbf{no-assessments (4)}

\item\textbf{difficult-to-cheat (4)}

\item\textbf{proctored-exam (4)}

\emph{by-a-friend (1)}

\item\textbf{take-home-exams (4)}

\item\textbf{covid-safety (4)}

\item\textbf{zoom (4)}

\emph{video-on (3), mic-on (1), polls (1)}

\item\textbf{prefers-in-person (4)}

\item\textbf{unconcerned (3)}

\item\textbf{test-versions (3)}

\item\textbf{tools-vary (3)}

\item\textbf{inconsistent-proctor-reports (3)}

\emph{reports-of-videos (2)}

\item\textbf{control-computer (3)}

\emph{restrict-activities (2)}

\item\textbf{test-integrity (3)}

\item\textbf{next-best-option (3)}

\item\textbf{open-note (3)}

\item\textbf{only-option-remotely (3)}

\item\textbf{required (3)}

\item\textbf{ensures-fairness (3)}

\item\textbf{enforce-exam-rules (3)}

\item\textbf{concern-with-camera (3)}

\item\textbf{technology-issues (3)}

\emph{recording (2), low-bandwidth (1), no-verifying-technology-works-before-exam (1)}

\item\textbf{checked-individual-assignments (3)}

\item\textbf{essay-based-exam (3)}

\item\textbf{covid (3)}

\emph{absence (1), infection-risk (1)}

\item\textbf{oral-exams (3)}

\item\textbf{badly (3)}

\item\textbf{unclear (3)}

\item\textbf{same-as-previous (3)}

\item\textbf{does-not-use (3)}

\item\textbf{recommended (3)}

\emph{respondus-lockdown-browser (1), zoom-room (1), by-school (1)}

\item\textbf{limits-academic-dishonesty (2)}

\item\textbf{academic-integrity (2)}

\emph{test-integrity (1)}

\item\textbf{remove-monitoring (2)}

\emph{webcam (1), face-detection (1)}

\item\textbf{no-benefits (2)}

\item\textbf{easy-to-implement (2)}

\item\textbf{saves-faculty-time (2)}

\item\textbf{implies-lack-of-trust (2)}

\item\textbf{system-is-limited (2)}

\item\textbf{outsource-proctoring (2)}

\item\textbf{would-not-use-any-monitoring-methods (2)}

\item\textbf{exam-integrity (2)}

\item\textbf{not-required (2)}

\item\textbf{self-proctoring (2)}

\emph{zoom (1), can-not-identify-sophisticated-cheating (1)}

\item\textbf{used-previously (2)}

\item\textbf{same-as-before (2)}

\item\textbf{used-default-methods (2)}

\item\textbf{student-discomfort (2)}

\emph{monitoring (1)}

\item\textbf{unethical (2)}

\emph{malware (1)}

\item\textbf{online-courses (2)}

\item\textbf{unreliable (2)}

\item\textbf{used-more-space (2)}

\item\textbf{security (2)}

\item\textbf{student-accessibility (2)}

\emph{what-worked-on-their-computer (1)}

\item\textbf{fairness (2)}

\item\textbf{cost (2)}

\item\textbf{ensures-integrity (2)}

\item\textbf{convenience (1)}

\item\textbf{students-select-enviroment (1)}

\item\textbf{professor-not-in-control (1)}

\item\textbf{easily-available (1)}

\emph{business-school (1)}

\item\textbf{wants-better-info-from-university (1)}

\item\textbf{questions-during-exam-difficult (1)}

\item\textbf{open-internet (1)}

\item\textbf{verify-student-id (1)}

\item\textbf{no-answer-keys (1)}

\item\textbf{avoids-technology-problems (1)}

\emph{providing-links-for-browser-downloads (1), practice-quizzes (1)}

\item\textbf{not-feasible (1)}

\item\textbf{control-testing-environment (1)}

\item\textbf{monitor-computer (1)}

\emph{computer-inputs (1), other-software (1)}

\item\textbf{would-use-recording (1)}

\item\textbf{completion-grading (1)}

\item\textbf{stay-on-schedule (1)}

\item\textbf{tradeoff (1)}

\emph{equity-versus-privacy (1), additional-stress (1)}

\item\textbf{clinicals (1)}

\item\textbf{popularity (1)}

\item\textbf{self-proctored (1)}

\item\textbf{used-lockdown (1)}

\item\textbf{accountability (1)}

\item\textbf{portfolio-show (1)}

\item\textbf{exam-type (1)}

\item\textbf{lax-on-cheating (1)}

\item\textbf{lack-of-trust (1)}

\item\textbf{exam-seriousness (1)}

\item\textbf{exam-questions-posted-online (1)}

\emph{chegg (1)}

\item\textbf{cheating-did-not-help (1)}

\item\textbf{add-monitoring (1)}

\emph{copy-paste-restrictions (1)}

\item\textbf{lack-honesty (1)}

\item\textbf{would-not-use-video (1)}

\item\textbf{higher-level-of-proctoring (1)}

\item\textbf{limit-student-stress (1)}

\item\textbf{can-be-avoided (1)}

\item\textbf{ease-of-grading (1)}

\item\textbf{students-designed-rubric (1)}

\item\textbf{stability (1)}

\item\textbf{colleagues-used-it (1)}

\emph{respondus (1)}

\item\textbf{wants-feedback-from-recorded-sessions (1)}

\item\textbf{uncomfortable (1)}

\emph{used-specifically-for-cheating (1)}

\item\textbf{department-requires (1)}

\item\textbf{harder-tests (1)}

\item\textbf{no-answer (1)}

\item\textbf{could-not-use-camera (1)}

\emph{students-uncomfortable (1)}

\item\textbf{false-positives (1)}

\item\textbf{AI-magic (1)}

\item\textbf{standerized-tests (1)}

\item\textbf{used-what-they-could (1)}

\item\textbf{can-be-done-online (1)}

\item\textbf{test-fairness (1)}

\item\textbf{SON-policy (1)}

\item\textbf{well (1)}

\item\textbf{written-language-proficiency (1)}

\item\textbf{provide-record-of-student-behavior (1)}

\item\textbf{would-modify-exam (1)}

\emph{does-not-need-proctoring (1)}

\item\textbf{test-in-class (1)}

\emph{student-preference (1)}

\item\textbf{RP-exams (1)}

\item\textbf{university-should-train-faculty (1)}

\item\textbf{grade-confidence (1)}

\item\textbf{simulate-in-class-exam (1)}

\item\textbf{async-exams (1)}

\item\textbf{inaccurate-reporting (1)}

\emph{caused-lockdown (1)}

\item\textbf{one-on-one-exam (1)}

\item\textbf{asks-students (1)}

\item\textbf{poor-integration-blackboard (1)}

\item\textbf{harms-learning (1)}

\item\textbf{special-arrangements (1)}

\item\textbf{enforce-time-limit (1)}

\item\textbf{would-use-lockdown (1)}

\item\textbf{not-applicable (1)}

\item\textbf{familiarity (1)}

\item\textbf{methods-not-necessary (1)}

\item\textbf{test-efficiency (1)}

\item\textbf{students-should-obey-us (1)}

\item\textbf{remain-on-schedule (1)}

\item\textbf{presentations (1)}

\item\textbf{prefer-monitor-in-person (1)}

\item\textbf{smaller-assignments (1)}

\item\textbf{not-allowed (1)}

\item\textbf{off-site-exam (1)}

\item\textbf{login-process (1)}

\item\textbf{performance-based-exam (1)}

\item\textbf{is-necessary-tradeoff (1)}

\item\textbf{proctored-exam-virtually (1)}

\item\textbf{uncomfortable-watching-recordings (1)}

\emph{student-living-situation (1)}

\item\textbf{no-choice-of-methods (1)}

\item\textbf{would-keep-settings (1)}

\item\textbf{misunderstood (1)}

\item\textbf{university-should-select-monitoring-system (1)}

\item\textbf{consult-outside-sources (1)}

\emph{unhelpful (1)}

\item\textbf{lecture-format (1)}

\item\textbf{resubmissions-allowed (1)}

\item\textbf{have-to-review-recordings (1)}

\item\textbf{increased-bandwidth (1)}

\item\textbf{turn-it-in (1)}

\item\textbf{convenient (1)}

\item\textbf{pandemic-safety (1)}

\item\textbf{prefer-live-monitoring (1)}

\item\textbf{not-enough-staff-for-watching (1)}

\item\textbf{allows-monitoring (1)}

\item\textbf{scientific (1)}

\item\textbf{distractions (1)}

\emph{roommates (1)}

\item\textbf{timed-exam-remotely (1)}

\item\textbf{exam-over-days (1)}

\emph{caught-cheating (1)}

\item\textbf{could-use-audio-files-in-test (1)}

\item\textbf{fully-online-class (1)}

\item\textbf{not-paid-enough-to-learn-systems (1)}

\item\textbf{lockdown-browser (1)}

\item\textbf{pop-quizzes (1)}

\item\textbf{students-select-time (1)}

\item\textbf{response-time (1)}

\emph{slow (1)}

\item\textbf{controlled-testing-environment (1)}

\emph{during-virtual (1)}

\item\textbf{faculty-mistrust-of-product (1)}

\end{itemize}

\end{document}